\documentclass[a4paper,11pt]{article}
\pdfoutput=1

\usepackage{jheppub} 
\usepackage{natbib}
\usepackage[utf8]{inputenc}

\usepackage{hyperref}
\usepackage{bbold}

\usepackage{graphicx}
\usepackage{amsmath,amsfonts,amssymb,amsbsy}
\usepackage{mathtools}
\usepackage{physics}

\usepackage{hyperref}
\usepackage{xcolor}
\hypersetup{
    colorlinks=true,
    allcolors={blue!60!black}
}

\usepackage{multirow}
\usepackage{stackrel}
\usepackage{tikz,pgfplots,pgfmath}
\usetikzlibrary{arrows.meta,decorations.markings,calc,intersections}

\usepackage{ifdraft}
\usepackage[]{easy-todo}

\usepackage{placeins}
\usepackage{slashed} 

\usepackage{booktabs}
\usepackage{adjustbox}
\usepackage{array}

\usepackage{cleveref}

\usepackage{subcaption}

\newcommand{\eps}{\epsilon}

\def\NC{{N_c}}
\def\NF{{N_f}}

\newcommand\Caravel{{\textsc{Caravel}}}

\def\trFive{{\rm tr}_5}

\newcommand{\eqnDiag}[1]{ \vcenter{\hbox{ #1}} }

\newcommand{\ii}{\,\mathrm{i}\,}

\begin{document}

\title{
    Leading-Color Two-Loop QCD Corrections for Three-Photon Production at Hadron Colliders
}
\preprint{{\footnotesize CERN-TH-2020-176, FR-PHENO-2020-016, MPP-2020-190}}

\author[a,b]{S.~Abreu,}
\author[c]{B.~Page,}
\author[d]{E.~Pascual,}
\author[e]{and V.~Sotnikov}

\emailAdd{samuel.abreu@cern.ch}
\emailAdd{bpage@ipht.fr}
\emailAdd{evgenij.pascual@saturn.uni-freiburg.de}
\emailAdd{sotnikov@mpp.mpg.de}

\affiliation[a]{Theoretical Physics Department, CERN, 1211 Geneva 23, Switzerland}
\affiliation[b]{Mani L.~Bhaumik Institute for Theoretical Physics,
Department of Physics and Astronomy, UCLA, Los Angeles, CA 90095, USA}
\affiliation[c]{Laboratoire  de  physique  de  l’Ecole  normale  sup\'erieure,
  ENS, Universit\'e  PSL,  CNRS,  Sorbonne  Universit\'e,  Universit\'e
  Paris-Diderot, Sorbonne  Paris  Cit\'e,  24  rue  Lhomond,  75005  Paris,  France}
\affiliation[d]{Physikalisches Institut,
Albert-Ludwigs-Universit\"at Freiburg, Hermann-Herder-Str.~3, 
D-79104 Freiburg, Germany}
\affiliation[e]{Max Planck Institute for Physics (Werner Heisenberg Institute),
D--80805 Munich, Germany}

\abstract{
We compute the two-loop helicity amplitudes for the production of three
photons at hadron colliders in QCD at leading-color.
Using the two-loop numerical unitarity method coupled with analytic reconstruction 
techniques, we obtain the decomposition of the two-loop 
amplitudes in terms of master integrals in analytic form. 
These expressions are valid to all orders in the dimensional regulator.
We use them to compute the two-loop finite remainders, which are 
given in a form that can be efficiently evaluated across the whole physical phase space. 
We further package these results in a public code which assembles the 
helicity-summed squared two-loop remainders, 
whose numerical stability across phase-space is demonstrated.
This is the first time that a five-point two-loop process is publicly available
for immediate phenomenological applications.
}

\maketitle



\section{Introduction}\label{sec:intro}

Precise theoretical predictions for scattering experiments at particle colliders
crucially rely on the availability of higher-order scattering amplitudes. Over
the next few years, the large increase in the amount of data collected by the
experiments at the Large Hadron Collider (LHC) at CERN will translate into
measurements made at unprecedented accuracy. To make the most of the
physics program of the LHC it is fundamental that theoretical predictions 
reach comparable levels of precision.
A large number of  $2\to2$ scattering processes are now
available including next-to-next-to-leading (NNLO) order QCD corrections
(see e.g.\ \cite{Amoroso:2020lgh} for a recent review).
However, reaching the same level of precision for higher multiplicity 
processes still remains a formidable challenge, with the best available
predictions only including next-to-leading order (NLO)  QCD  and electro-weak
corrections.

In this work, we focus on the double-virtual NNLO QCD corrections to the production 
of three photons at hadron colliders. This process is important for studying various 
beyond-the-Standard-Model (BSM) phenomena. In particular, it can be used
to constrain anomalous quartic gauge~\cite{Aad:2015bua} and Higgs 
couplings~\cite{Denizli:2019ijf,Denizli:2020wvn,Aguilar-Saavedra:2020rgo}.
Furthermore, it is a major contribution to the irreducible background in searches 
for associated production of BSM particles and a photon~\cite{Zerwekh:2002ex,Toro:2012sv,Das:2015bda}. 
Similar to the related two-photon production 
process~\cite{Anastasiou:2002zn,Glover:2003cm,Catani:2011qz,Campbell:2016yrh,Gehrmann:2020oec,Alioli:2020qrd},
three-photon production exhibits very slow perturbative convergence~\cite{Chawdhry:2019bji,Kallweit:2020gcp} and
non-reliable estimates of uncertainty from missing higher orders.
In particular, a significant tension between LHC Run 1 data and NLO QCD predictions has been reported~\cite{Aaboud:2017lxm}.
The computation of NNLO QCD corrections is thus crucial for this process.
While several NNLO infrared subtraction approaches are
capable of tackling triphoton production \cite{Chawdhry:2019bji,Kallweit:2020gcp}, 
obtaining the two-loop five-point amplitudes required for the double-virtual contributions
in a representation that allows for an efficient on-the-fly computation
remains a major obstacle. Our goal is to provide such
expressions, and by doing so making these amplitudes widely available for any future
phenomenological studies involving this process.

In the last few years, there has been remarkable progress in the calculation of 
two-loop five-point amplitudes.
A basis of master integrals relevant for the scattering of five massless particles
has been known for a number of years in the planar 
limit~\cite{Papadopoulos:2015jft,Gehrmann:2015bfy}, 
and more recently including non-planar topologies~\cite{Abreu:2018aqd,Chicherin:2018old}.
These master integrals evaluate to a class of multi-valued special functions 
with logarithmic branch cuts.
Given the large number of scales in five-point kinematics, it is essential to find a representation
of the master integrals in terms of special functions that manifests the physical properties of amplitudes.
This was achieved by defining a set of so-called pentagon functions, first
for the planar integral topologies~\cite{Gehrmann:2018yef}, 
and more recently for a complete set of five-point massless master integrals~\cite{Chicherin:2020oor}.
The latter work also provides an efficient code for their numerical evaluation across all
of phase space.
At the same time, there has been substantial progress in the reduction of five-point two-loop scattering amplitudes to master 
integrals~\cite{Larsen:2015ped,Ita:2015tya,Peraro:2016wsq,Georgoudis:2016wff,Georgoudis:2017iza,
Chawdhry:2018awn,Boehm:2018fpv,Peraro:2019svx,Peraro:2019cjj,Bendle:2019csk,Wang:2019mnn,Guan:2019bcx,Abreu:2020xvt},
often building on the use of finite-field arithmetic~\cite{vonManteuffel:2014ixa,Peraro:2016wsq}, 
and functional reconstruction techniques~\cite{Peraro:2016wsq,Klappert:2019emp,Peraro:2019svx,Klappert:2020aqs,Klappert:2020nbg}.
As a result, all five-parton two-loop planar QCD amplitudes were computed 
numerically~\cite{Badger:2013gxa,Badger:2017jhb,Abreu:2017hqn, Badger:2018gip,Abreu:2018jgq}
and in analytic form~\cite{Badger:2018enw,Abreu:2018zmy,Badger:2019djh,Abreu:2019odu}.
Despite the remarkable progress of the last few years, five-point amplitudes 
are not yet broadly available in a form suitable for direct phenomenological applications.
Indeed, the only phenomenological study involving five-point two-loop amplitudes
is the work of~\cite{Chawdhry:2019bji}, where the authors evaluated planar 
two-loop $q\bar{q}\to \gamma\gamma\gamma$ amplitudes on a sufficient set of phase-space points 
to construct interpolating functions.

In this paper, we calculate a complete set of independent planar two-loop helicity amplitudes required 
for the double-virtual NNLO QCD corrections to three-photon production at hadron colliders. 
We present them for the first time in a form which is suitable for direct and flexible 
phenomenological applications. Indeed, our results have already been employed 
in~ref.~\cite{Kallweit:2020gcp} to implement the first on-the-fly computation of NNLO QCD corrections 
to triphoton production within the \textsc{Matrix} framework \cite{Grazzini:2017mhc}. We obtain the two-loop helicity amplitudes by following a similar approach to that used in
refs.~\cite{Abreu:2018zmy,Abreu:2019odu}, based on the two-loop numerical unitarity 
approach~\cite{Abreu:2017xsl,Abreu:2017hqn} as implemented in the recently released 
\texttt{C++} library \Caravel{}~\cite{Abreu:2020xvt}.
Unlike in refs.~\cite{Abreu:2018zmy,Abreu:2019odu}, where only the finite remainders were obtained, in this paper we
obtain analytic expressions for the decomposition of the amplitudes into
master integrals, marking the first time that such results are available for
five-point two-loop amplitudes. This decomposition is valid to all orders in the dimensional regulator.
They are valuable for studying the analytic complexity of the master-integral decomposition and for
future computations of higher-order corrections.
We also obtain analytic expressions for the two-loop finite remainders, 
decomposed into the pentagon functions of ref.~\cite{Chicherin:2020oor}.
We confirm that the latter are more suitable for efficient and stable numerical evaluations.
All the analytic results we obtain are made available in a set of ancillary files, 
and we provide a public \texttt{C++} library~\cite{FivePointAmplitudes} for the efficient
numerical evaluation of the finite remainders.

The paper is organized as follows. In \cref{sec:notAndConv}
we establish our conventions and define the objects we will be computing.
In \cref{sec:remainders}, we present our approach to the calculation of
the two-loop amplitudes and their finite remainders.
We discuss some properties of the analytic results we obtained, 
give reference evaluation values and discuss the checks that were performed.
In \cref{sec:results}, we present a public implementation of the numerical evaluation of the analytic results for finite remainders
and demonstrate its efficiency and numerical stability.


\section{Notation and Conventions}\label{sec:notAndConv}

\subsection{Helicity Amplitudes}

We consider a complete set of helicity amplitudes required for the 
computation of the double-virtual next-to-next-to leading order (NNLO) 
QCD corrections to the production of 
three photons at hadron colliders.
More precisely, we consider the parton-level scattering process%
\begin{equation}\label{eq:process}
  q(p_1,h_1)+\bar{q}(p_2,h_2)\,\to\,\gamma(p_3,h_3)+\gamma(p_4,h_4)+\gamma(p_5,h_5)\,.
\end{equation}
This is the only (sub-)process required for these corrections, as the loop-induced process 
$gg\to \gamma\gamma\gamma$ vanishes to all orders in the coupling constants
due to the charge conjugation symmetry of QCD$\otimes$QED.
We denote the momenta and the helicity states of the particles as $p_i$ and $h_i$ respectively. 
All particles are massless, and the kinematics of the process are thus fully specified by the
five independent Mandelstam invariants
\begin{align}\begin{split}\label{eq:mandDef}
  s_{12}=(p_1+&p_2)^2\,,\quad
  s_{23}=(p_2+p_3)^2\,,\quad
  s_{34}=(p_3+p_4)^2\,,\quad\\
  &s_{45}=(p_4+p_5)^2\,,\quad
  s_{15}=(p_1+p_5)^2\,,\quad
\end{split}\end{align}
and the parity-odd contraction of four momenta
\begin{equation}\label{eq:tr5}
  \trFive \coloneqq 4\ii\varepsilon(p_1,p_2,p_3,p_4),
\end{equation}
where $\varepsilon(\boldsymbol\cdot,\boldsymbol\cdot,\boldsymbol\cdot,\boldsymbol\cdot )$ is the fully anti-symmetric Levi-Civita symbol.
The physical region associated with the process in \cref{eq:process},
where momenta $ p_1$ and $ p_2$ are incoming, is characterized by
\begin{equation}\label{eq:physregion}
  s_{12},s_{34},s_{45} > 0, \qquad s_{23},s_{15} < 0, \qquad \trFive^2 < 0.
\end{equation}
The latter condition is equivalent to the negativity of the five-point Gram 
determinant and it is trivially satisfied by real-valued momenta in \cref{eq:tr5}.

The amplitudes for this process, denoted as 
$\mathcal{M}(1_q^{h_1},2_{\bar{q}}^{h_2},3_\gamma^{h_3},4_\gamma^{h_4},5_\gamma^{h_5})$,
can be decomposed into a color factor, a helicity-dependent spinor weight, 
and a Lorentz invariant kinematic factor. 
That is, we can write
\begin{equation}\label{eq:def_A}
  \mathcal{M}(1_q^{h_1},2_{\bar{q}}^{h_2},3_\gamma^{h_3},4_\gamma^{h_4},5_\gamma^{h_5})
  \coloneqq e_q^3\delta_{i_1i_2}
  \Phi(1_q^{h_1},2_{\bar{q}}^{h_2},3_\gamma^{h_3},4_\gamma^{h_4},5_\gamma^{h_5})
  \mathcal{A}(1_q^{h_1},2_{\bar{q}}^{h_2},3_\gamma^{h_3},4_\gamma^{h_4},5_\gamma^{h_5})\,,
\end{equation}
where $i_1$ and $i_2$ are color indices of the quark and the antiquark, 
$e_q$ is the electric charge of the external quark in the process, and $\Phi$ denotes the 
spinor-weight factor.
In the following, we will call 
$\mathcal{A}(1_q^{h_1},2_{\bar{q}}^{h_2},3_\gamma^{h_3},4_\gamma^{h_4},5_\gamma^{h_5})$
the \emph{helicity amplitudes} for the process in eq.~\eqref{eq:process}. 
For simplicity, we will often suppress the arguments of $\mathcal{A}$.
We employ the 't Hooft-Veltman scheme of dimensional regularization with 
$D=4-2\epsilon$ space-time dimensions to regularize infrared and ultraviolet 
divergences.
We define the dimensionally regularized helicity amplitudes 
with external quarks as in \cite{Abreu:2018jgq}.

The bare helicity amplitudes have a perturbative expansion in powers of the bare strong coupling $\alpha^{0}_{s}=(g^{0}_{s})^2/(4\pi)$, which we write as
\begin{equation}
  \mathcal{A} = \mathcal{A}^{(0)}+\frac{\alpha^{0}_{s}}{2\pi}\mathcal{A}^{(1)}
  +\left(\frac{\alpha^{0}_{s}}{2\pi}\right)^2\mathcal{A}^{(2)}
  +\mathcal{O}\left((\alpha^{0}_s)^3\right)\,.
\end{equation}
The renormalized coupling $\alpha_s$ is related to the bare $\alpha_s^0$ through
\begin{equation}\label{eq:renormCoupling}
  \alpha_s^{0}\mu_0^{2\epsilon}S_{\epsilon} =
    \alpha_s\mu^{2\epsilon}\left( 1-\frac{\beta_0}{\epsilon}\frac{\alpha_s}{2\pi}+ \mathcal{O} \left(\alpha_s^2\right)\right), \qquad S_\epsilon=(4\pi)^{\eps}e^{-\eps\gamma_E},
\end{equation}
where $\gamma_E$ is the Euler-Mascheroni constant, and $\mu_0$ and $\mu$ are the 
dimensional regularization and renormalization scales, which from now on we assume to be equal.
$\beta_0$ is the first coefficient of the QCD $\beta$-function,
\begin{equation}
  \beta_0=\frac{11 C_A - 4 T_F \NF }{6} , 
\end{equation}
where $C_A=\NC$ is the quadratic Casimir of the adjoint representation of the $SU(\NC)$ group,
and $T_F = 1/2$ is the normalization of fundamental representation generators. Below we will also
need the quadratic Casimir of the fundamental representation, $C_F = \frac{\NC^2-1}{2\NC}$.
We define the perturbative expansion of the renormalized amplitudes as
\begin{equation}\label{eq:renormAmpExp}
  \mathcal{A}_R=\mathcal{A}^{(0)}_R+\frac{\alpha_s}{2\pi}\mathcal{A}^{(1)}_R
  +\left(\frac{\alpha_s}{2\pi}\right)^2\mathcal{A}^{(2)}_R+\mathcal{O}(\alpha_s^3)\,.
\end{equation}
The coefficients $\mathcal{A}^{(k)}_R$ are then related to their bare counterparts as
\begin{align} \label{eq:twoLoopUnRenorm}
  \mathcal{A}_R^{(0)}=\mathcal{A}^{(0)}\,, \quad
  \mathcal{A}_R^{(1)}=S_{\epsilon}^{-1}\mathcal{A}^{(1)}\,,\quad
  \mathcal{A}_R^{(2)}=
  S_{\epsilon}^{-2}\mathcal{A}^{(2)}
  -\frac{\beta_0}{\epsilon}  S_{\epsilon}^{-1}
  \mathcal{A}^{(1)}\,.
\end{align}

There are 16 different helicity configurations to consider. However, it can 
be easily shown that only 2 of them are independent. 
In this paper we choose
the independent configurations to be
\begin{equation}\label{eq:independentHel}
  \begin{aligned}
    \mathcal{A}_{+++}(1,2,3,4,5) &\coloneqq \mathcal{A}(1_q^+,2_{\bar{q}}^-,3_\gamma^+,4_\gamma^+,5_\gamma^+)\,,\\
    \qquad
    \mathcal{A}_{-++}(1,2,3,4,5) &\coloneqq \mathcal{A}(1_q^+,2_{\bar{q}}^-,3_\gamma^-,4_\gamma^+,5_\gamma^+)\,,
  \end{aligned}
\end{equation}
where we indexed each independent amplitude by the photon helicities.
In table \ref{tab:hel_rel}, we show how permutations of momenta, parity and charge conjugation can be used to relate all helicity configurations to these two amplitudes.
Note that all momentum permutations are within the scattering region defined in \cref{eq:physregion}.
We choose $\Phi_{-++}$ such that $\mathcal{A}_{-++}^{(0)} = 1$ and we choose
\begin{equation}
  \Phi_{+++} = \frac{[3 1] \langle 1 2\rangle ^3 \langle 1 3\rangle }{\langle 1 4\rangle ^2 \langle 1 5\rangle ^2 \langle 2 3\rangle ^2}\,.
\end{equation}

\begin{table}
  \begin{minipage}{0.5\textwidth}
    \centering
    \begin{tabular}{cc}
      \toprule
      Helicity & Expression \\
      \midrule
      $+ -~+ + +$ &  $\mathcal{A}_{+++}(1,2,3,4,5)$ \\
      $+ -~- + +$ &  $\mathcal{A}_{-++}(1,2,3,4,5)$ \\
      $+ -~+ - +$ &  $\mathcal{A}_{-++}(1,2,4,3,5)$ \\
      $+ -~+ + -$ &  $\mathcal{A}_{-++}(1,2,5,3,4)$ \\

      $+ -~- - +$ &  $\mathbf{P} \mathcal{A}_{-++}(2,1,5,3,4)$ \\
      $+ -~+ - -$ &  $\mathbf{P} \mathcal{A}_{-++}(2,1,3,4,5)$ \\
      $+ -~- + -$ &  $\mathbf{P} \mathcal{A}_{-++}(2,1,4,3,5)$ \\
      $+ -~- - -$ &  $\mathbf{P} \mathcal{A}_{+++}(2,1,3,4,5)$ \\
      \bottomrule
    \end{tabular}
  \end{minipage}
  \begin{minipage}{0.5\textwidth}
    \centering
    \begin{tabular}{cc}
      \toprule
      Helicity & Expression \\
      \midrule
      $- +~+ + +$ &  $\mathcal{A}_{+++}(2,1,3,4,5)$ \\
      $- +~- + +$ &  $\mathcal{A}_{-++}(2,1,3,4,5)$ \\
      $- +~+ - +$ &  $\mathcal{A}_{-++}(2,1,4,3,5)$ \\
      $- +~+ + -$ &  $\mathcal{A}_{-++}(2,1,5,3,4)$ \\

      $- +~- - +$ &  $\mathbf{P} \mathcal{A}_{-++}(1,2,5,3,4)$ \\
      $- +~+ - -$ &  $\mathbf{P} \mathcal{A}_{-++}(1,2,3,4,5)$ \\
      $- +~- + -$ &  $\mathbf{P} \mathcal{A}_{-++}(1,2,4,3,5)$ \\
      $- +~- - -$ &  $\mathbf{P} \mathcal{A}_{+++}(1,2,3,4,5)$ \\
      \bottomrule
    \end{tabular}
  \end{minipage}  
  \caption{
    Relation between all helicity configurations and
    the two independent basis elements in eq.~\eqref{eq:independentHel}.
    $\mathbf{P}$ denotes a parity transformation.
  }
  \label{tab:hel_rel}
\end{table}

\begin{figure}
  \centering
  \begin{minipage}{0.45\textwidth}\centering
    \includegraphics[scale=0.6]{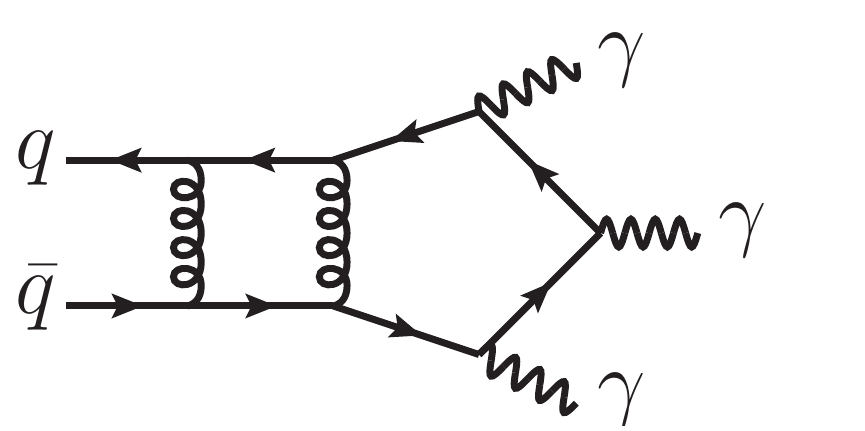}
    \subcaption{Contribution proportional to $C_F^2$}
    \label{fig:cf2}
  \end{minipage}
  \begin{minipage}{0.45\textwidth}\centering
    \includegraphics[scale=0.6]{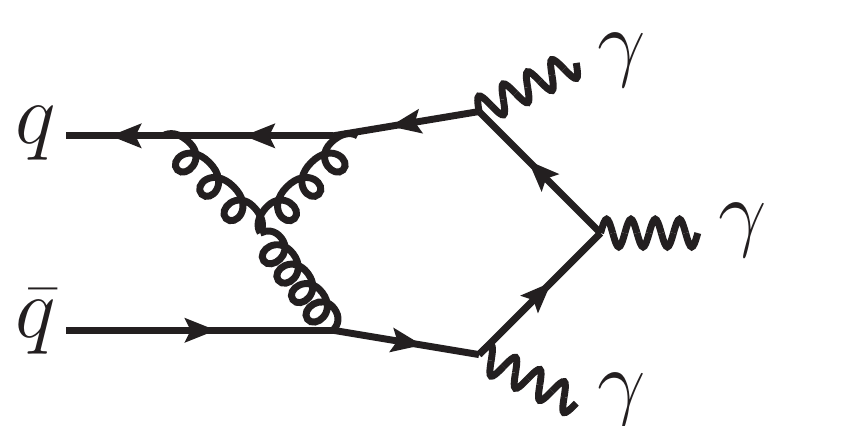}
    \subcaption{Contribution proportional to $C_FC_A$}
    \label{fig:cfca}
  \end{minipage}\vspace{3mm}
  \begin{minipage}{0.45\textwidth}\centering
    \includegraphics[scale=0.6]{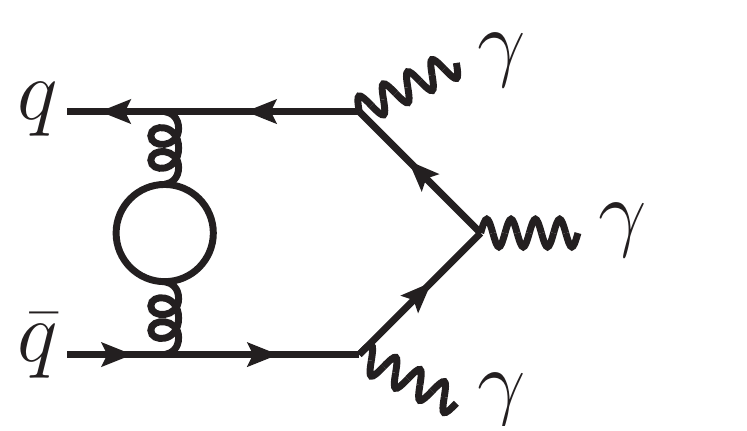}
    \subcaption{Contribution proportional to $N_f$}
    \label{fig:nf}
  \end{minipage}
  \begin{minipage}{0.45\textwidth}\centering
    \includegraphics[scale=0.6]{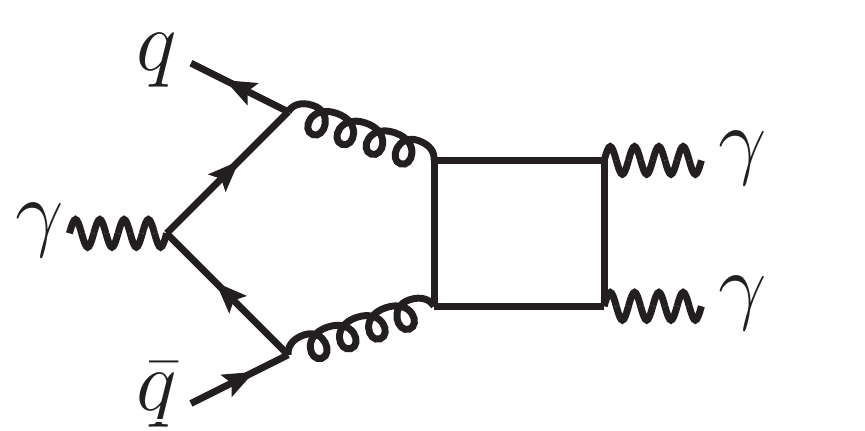}
    \subcaption{Contribution proportional to $\sum_f {Q^2_f}$}
    \label{fig:sumQ}
  \end{minipage}
  \caption{Representative diagrams for each of the contributions in 
  eq.~\eqref{eq:coulourDec}. Photons ($\gamma$) are denoted by wavy lines,
  gluon by curly lines, and quarks by straight lines.}
  \label{fig:sampleDiag}
\end{figure}

Each helicity amplitude can be further decomposed into individually gauge-invariant contributions as
\begin{align} \begin{split}\label{eq:coulourDec}
    \mathcal{A}^{(1)} & = C_F A^{(1)}, \\
    \mathcal{A}^{(2)} & = C_F^2 B^{(2,0)} + C_F C_A B^{(2,1)} + 
    C_F T_F \NF A^{(2,\NF)} + C_F T_F \left(\sum_{f=1}^{\NF} Q_f^2\right) \, \tilde{A}^{(2,N_f)},
\end{split}\end{align}
where $\NF$ is the number of light quarks,
and $Q_f$ is the ratio of the electric charges of the quark 
with flavor $f$ and the quark in the initial state.
Here we do not consider the contributions from heavy quark loops.
In \cref{fig:sampleDiag} we depict representative diagrams for each of these contributions.\footnote{
  Contributions similar to those in fig.~\ref{fig:sumQ} but where either a single
  or three photons attach to the closed quark loop vanish due to 
  charge-conjugation symmetry of QED$\otimes$QCD.
}

In this paper, we compute the two-loop amplitudes $\mathcal{A}^{(2)}$ in the 
leading-color approximation, where the number of colors $N_c$ is large with 
the ratio $\NF/\NC$ kept constant.
In this approximation only planar topologies contribute.
We also do not consider the gauge-invariant 
term $\tilde{A}^{(2,N_f)}$ which includes non-planar contributions.
The validity of this approximation for phenomenology is discussed in ref.~\cite{Kallweit:2020gcp}.
We write the two-loop amplitudes~$\mathcal{A}^{(2)}$ as
\begin{align}\label{eq:leadingColourDec}
  \mathcal{A}^{(2)} = \frac{N_c^2}{4}\left(A^{(2,0)} + \mathcal{O}(\NC^{-2})  \right)  + C_F T_F \NF A^{(2,\NF)}, \qquad A^{(2,0)} = B^{(2,0)} + 2 B^{(2,1)}.
\end{align}
In summary, the main goal of this paper consists in the calculation of
four functions: $A^{(2,0)}_{+++}$, $A^{(2,\NF)}_{+++}$,
$A^{(2,0)}_{-++}$ and $A^{(2,\NF)}_{-++}$. 
We note that since $\mathcal{A}_{+++}^{(0)}=0$, only the latter
two are required to compute NNLO QCD corrections 
to the process in \cref{eq:process}.

\subsection{Finite Remainders}\label{sec:finiteRemainders}

At NNLO, the two-loop amplitudes contribute to the cross section only through finite remainders (see e.g.~\cite{Weinzierl:2011uz}),
which can be expanded similarly to eq.~\eqref{eq:renormAmpExp} as
\begin{equation}
  \mathcal{R} = \mathcal{R}^{(0)} + \frac{\alpha_s}{2\pi} \mathcal{R}^{(1)} + \left(\frac{\alpha_s}{2\pi}\right)^2 \mathcal{R}^{(2)} + \mathcal{O}(\alpha_s^3)\,.
\end{equation}
The coefficients $\mathcal{R}^{(i)}$ are obtained from the expansion of the renormalized amplitudes
$\mathcal{A}_R$ by subtracting the remaining infrared singularities \cite{Catani:1998bh}.
More precisely, we define
\begin{align}\begin{split} \label{eq:remainder2l}
    \mathcal{R}^{(0)} &= \mathcal{A}_{R}^{(0)}, \\
    \mathcal{R}^{(1)} &= \mathcal{A}_R^{(1)}-\mathbf{I}^{(1)}\mathcal{A}_R^{(0)} ~+\mathcal{O}(\epsilon), \\ 
    \mathcal{R}^{(2)} &= \mathcal{A}_R^{(2)}-\mathbf{I}^{(1)}\mathcal{A}_R^{(1)}-\mathbf{I}^{(2)}{\cal A}_R^{(0)} ~+\mathcal{O}(\epsilon),
\end{split}\end{align}
where the functions $\mathbf{I}^{(1)}$ and $\mathbf{I}^{(2)}$ are defined as 
(see e.g.~\cite{Anastasiou:2002zn,Glover:2003cm})
\begin{align}\begin{split}\label{eq:catanii}
\mathbf{I}^{(1)} &(\epsilon)  = - C_F \frac{e^{\gamma_E\epsilon}}{\Gamma(1-\epsilon)}\left(\frac{1}{\epsilon^2}+\frac{3}{2\epsilon}\right)\left(-\frac{s_{12}}{\mu^2}-\ii 0\right)^{-\epsilon}, \\
    \mathbf{I}^{(2)} &(\epsilon) = 
    -\frac{1}{2}\mathbf{I}^{(1)}(\epsilon)\mathbf{I}^{(1)}(\epsilon)-\frac{\beta_0}{\epsilon}\mathbf{I}^{(1)}(\epsilon)+\frac{e^{-\gamma_E\epsilon}\Gamma(1-2\epsilon)}{\Gamma(1-\epsilon)}\left(\frac{\beta_0}{\epsilon}+K\right)\mathbf{I}^{(1)}(2\epsilon)+\mathbf{H}(\epsilon).
\end{split}\end{align}
In $\mathbf{I}^{(2)}(\epsilon)$, we have introduced the functions
\begin{align}\begin{split}
    K & =\left(\frac{67}{18}-\frac{\pi^2}{6}\right) C_A - \frac{10}{9} T_F N_F, \qquad
    \mathbf{H}(\epsilon)=\frac{e^{\gamma_E\epsilon}}{2\epsilon\Gamma(1-\epsilon)} H_q, \\ 
    H_q  &= \left( \frac{\pi^2}{2} -6 \zeta_3 -\frac{3}{8}\right) C_F^2 + \left(\frac{13}{2} \zeta_3 + \frac{245}{216} - \frac{23}{48}\pi^2\right) C_A C_F +  \left( \frac{\pi^2}{12}-\frac{25}{54} \right) T_F C_F \NF.
\end{split}\end{align}
The finite
remainders can be decomposed similarly to \cref{eq:leadingColourDec}, and we
write
\begin{align}\begin{split}\label{eq:remaindersDef}
    \mathcal{R}^{(1)} &= C_F R^{(1)}, \\
    \mathcal{R}^{(2)} &= \frac{N_c^2}{4}\left(R^{(2,0)} + \mathcal{O}(\NC^{-2})  \right) + C_F T_F \NF R^{(2,\NF)}\,.
\end{split}\end{align}
Analogously to \cref{eq:independentHel}, whenever convenient we will also
write $R^{(i,j)}_{+++}$ and $R^{(i,j)}_{-++}$ to denote the contributions
to each of the two independent helicity states.


\section{Calculation of Helicity Amplitudes}\label{sec:remainders}

To compute the functions $A^{(2,0)}_{+++}$, $A^{(2,\NF)}_{+++}$,
$A^{(2,0)}_{-++}$ and $A^{(2,\NF)}_{-++}$ defined
in \cref{eq:leadingColourDec}, we use the framework of two-loop numerical 
unitarity~\cite{Ita:2015tya,Abreu:2017idw,Abreu:2017xsl,Abreu:2017hqn}
coupled with functional reconstruction techniques.
The same approach was already used previously to compute
the five-parton two-loop amplitudes \cite{Abreu:2018zmy,Abreu:2019odu}.
We build on the implementation of this framework in \Caravel~\cite{Abreu:2020xvt}, 
which we modify to handle amplitudes with external photons.
In this section, we summarize the main steps of our calculation.

Before delving into this, however, we note that our approach requires
knowing the one-loop amplitudes to order $\epsilon^2$. We have computed
them to all orders in $\epsilon$ using the same techniques as those used for the
two-loop amplitudes. This is by now an easy calculation so we will not discuss
it further, and simply include the results in ancillary files.

\subsection{Two-loop Numerical Unitarity}\label{sec:2lNumUni}

Our approach to the calculation of two-loop amplitudes is built on their numerical
evaluation within the framework of two-loop numerical unitarity. 
We target independently each of the helicity amplitudes $A^{(2,j)}_h$ with 
$j=\{0,\NF\}$, $h=\{+++,-++\}$,
see \cref{eq:leadingColourDec}.
The approach starts from a parametrization
of the integrand of the amplitude ${A}^{(2,j)}_h(\ell_l)$ in terms of
master integrands and surface terms \cite{Ita:2015tya} ($\ell_l$
denotes the loop momenta).
The master integrands are associated with master integrals, and the 
surface terms integrate to zero. This decomposition is naturally
organized in terms of propagator structures. More precisely, we write
\begin{equation}\label{eq:integrand}
  {A}^{(2,j)}_h(\ell_l)=\sum_{\Gamma\in\Delta}
  \sum_{i\in M_\Gamma\cup S_\Gamma}c_{\Gamma,i}
  \frac{m_{\Gamma,i}(\ell_l)}{\prod_{j\in P_\Gamma}\rho_j}\,,
\end{equation}
where $\Delta$ is the set of propagator structures $\Gamma$, $P_\Gamma$
is the multiset of inverse propagators $\rho_j$ in $\Gamma$, and $M_\Gamma$
and $S_\Gamma$ denote the sets of master integrands and surface terms. The coefficients 
$c_{\Gamma,i}$ are determined using the factorization properties of the integrand
${A}^{(2,j)}(\ell_l)$ in specific configurations $\ell_l^\Gamma$ of the loop
momenta where the inverse propagators $\rho_j\in P_\Gamma$ are on-shell, that
is $\rho_j(\ell_l^\Gamma) = 0$ iff $j\in P_\Gamma$. In this limit, the leading
contribution to eq.~\eqref{eq:integrand} factorizes as
\begin{equation}\label{eq:onshell}
  \sum_\text{states}\prod_{i\in T_\Gamma}{A}_i^{(0)}(\ell_l^\Gamma)=
  \sum_{\Gamma'\geq\Gamma,i\in M_{\Gamma'}\cup S_{\Gamma'}}
  \frac{c_{\Gamma',i}m_{\Gamma',i}(\ell_l^\Gamma)}
  {\prod_{j\in(P_{\Gamma'}\setminus P_\Gamma)}\rho_j(\ell_l^\Gamma)}.
\end{equation}
The sum on the right-hand side is over the propagator structures $\Gamma'$ 
such that $P_\Gamma\subseteq P_{\Gamma'}$. On the left-hand side, $T_\Gamma$ denotes
the set of tree amplitudes associated with the vertices in the diagram corresponding 
to $\Gamma$ and the sum is over the scheme-dependent physical states of each internal
line of $\Gamma$. 

We refer to \cref{eq:onshell} as \emph{cut equations}, and to its left hand side as \emph{cuts}.
The coefficients $c_{\Gamma,i}$ are determined by sampling the
cut equations over enough values of $\ell_l^\Gamma$.
To construct the color-stripped products of tree amplitudes on the left-hand side of \cref{eq:onshell},
we use the unitarity-based color decomposition approach of \cite{Ochirov:2016ewn, Ochirov:2019mtf} to include colorless particles.
The dimensional-regulator dependence of cuts is determined with the approach of 
decomposition by particle content \cite{Anger:2018ove,Abreu:2019odu,Sotnikov:2019onv},
based on dimensional reduction.
We evaluate the color-stripped and dimensional-regulator-free tree amplitudes through Berends-Giele recursion \cite{Berends:1987me}.
The resulting system of equations is then solved numerically on a
given phase-space point. All numerical operations are done
using finite-field arithmetic. This allows us to obtain exact coefficients 
$c_{\Gamma,i}$ for rational phase-space points.
The latter can be generated, for instance, by
using momentum-twistor variables \cite{Hodges:2009hk}.
Once the $c_{\Gamma,i}$ have been determined, we obtain the decomposition
of the amplitude in terms of master integrals,
\begin{equation}\label{eq:A}
  {A}^{(2,j)}_h =\sum_{\Gamma\in\Delta}
  \sum_{i\in M_\Gamma}c_{\Gamma,i}m_{\Gamma,i}\,,
\end{equation}
where $m_{\Gamma,i}$ is the master integral associated with the 
numerator $m_{\Gamma,i}(\ell_l)$.

Compared to five-parton amplitudes, we note that the fact that photons
cannot be ordered leads to a proliferation of topologies.
To be more explicit, consider as an example the two-loop
leading-color color-ordered five-gluon amplitudes.
Due to the color ordering, there are only 5 
different pentagon-box 
topologies\footnote{We call \emph{pentagon-box} the topology corresponding
  to the diagram shown in \cref{fig:cf2}.
} to consider, which 
can be labeled e.g.~by the vertex attached to the gluon of momentum
$p_1$. For the photon amplitudes we are considering in this
paper, we should multiply this number by $3!$, corresponding to 
the possible orderings of the photons. We thus have to consider
30 different pentagon-box topologies.

\subsection{Pentagon Functions}

A basis of master integrals for planar five-point massless amplitudes
is known \cite{Papadopoulos:2015jft,Gehrmann:2015bfy}. 
Order by order in $\epsilon$, they can be expressed in terms of
multiple polylogarithms. MPLs form a special class of functions
with logarithmic singularities, and this class of functions can be equipped with 
algebraic structures which allow one to find relations between 
them~\cite{Goncharov:2010jf,Duhr:2011zq,Duhr:2012fh}. As a consequence,
we can define a set of functions, called pentagon functions,
which form a basis for the MPLs that appear in the master integrals
contributing to planar two-loop five-point massless 
amplitudes~\cite{Gehrmann:2018yef,Chicherin:2020oor}.

After evaluating the coefficients $c_{\Gamma,i}$ in \cref{eq:integrand},
we directly obtain the decomposition of the helicity amplitudes in terms 
of master integrals, see \cref{eq:A}. 
Assuming that the decomposition of the master integrals 
in terms of pentagon functions is known, we in turn obtain a decomposition
of the amplitude in terms of pentagon functions. If we denote
the pentagon functions by $\{h_i\}_{i\in B}$, with $B$ the associated
set of labels, we can then decompose the amplitude as
\begin{equation}\label{eq:pent2l}
  {A}^{(2,j)}_h =\sum_{i\in B}\sum_{k=-4}^0\epsilon^k d_{k,i}h_i
  +\mathcal{O}(\epsilon)\,,
\end{equation}
where we make explicit that two-loop amplitudes have at most poles of order $\epsilon^{-4}$.
The decomposition of \cref{eq:pent2l} presents a major advantage 
compared to the decomposition of an amplitude in terms of master integrals:
it allows us to write one-loop and two-loop amplitudes in terms of a common
basis of functions. It then follows from \cref{eq:remainder2l,eq:remaindersDef}
that remainders themselves can be decomposed in terms of pentagon 
functions,\footnote{It is clear that the coefficients of the Laurent
expansion of $\mathbf{I}^{(1)}$ and $\mathbf{I}^{(2)}$ around $\epsilon=0$
also belong to this space of functions.}
that is,
\begin{equation}\label{eq:pentFunc}
  R^{(2,j)}_h =\sum_{i\in B}r_ih_i\,.
\end{equation}

In this paper, we will adopt the pentagon functions defined in ref.~\cite{Chicherin:2020oor}.
Aside form the fact that they can be efficiently evaluated across phase-space,
they are also defined to be a basis for the whole orbit under the 
symmetry group  of five-point kinematics in the $\{1,2\}$-channel (see \cref{eq:physregion}). 
This is crucial for our calculation. As already mentioned previously, the 
amplitudes receive contributions from all orderings of the photons
in the final state. An important consequence of this observation is that
there is no Euclidean region for the amplitude, that is no region
where the amplitude is real.

\subsection{Functional Reconstruction of Master Integral Coefficients}

In refs.~\cite{Abreu:2018zmy,Abreu:2019odu}, we argued that
having two-loop corrections in mind one should reconstruct
the analytic form of the coefficients $r_i$ of the decomposition of the remainders
in \cref{eq:pentFunc}. Indeed, these are expected to be much simpler than the
master integral coefficients in the decomposition of the amplitude in \cref{eq:A}.
To better understand the simplicity of the remainder in comparison to the two-loop
amplitudes for five-point QCD amplitudes, 
in this paper we reconstruct the coefficients
$c_{\Gamma,i}$ in \cref{eq:A}.
We thus obtain a decomposition for the amplitude that is valid to all orders in
$\epsilon$. This is the first time that such a decomposition has been made
available for a two-loop five-point QCD amplitude. While this expression
contains more information than required for the computation of NNLO corrections
to the process in \cref{eq:process}, it can be used for defining the three-loop
remainder required for N$^3$LO corrections. For NNLO applications, we can use it to
extract two-loop remainders in the form given in \cref{eq:pentFunc}.

Let us briefly discuss how the master-integral coefficients are computed.
We proceed following the same steps as in ref.~\cite{Abreu:2019odu},
which we adapt to the reconstruction of master-integral coefficients. 
For simplicity of the expressions, we use a single label to index
the master integrals and rewrite \cref{eq:A} as
\begin{equation}\label{eq:Abis}
{A}^{(2,j)}_h =\sum_{i}
  c_{i}(\epsilon,\vec s,\trFive)\,m_{i}(\epsilon,\vec s,\trFive)\,,
\end{equation}
where $\vec s$ denotes the set of five independent $s_{ij}$
defined in \cref{eq:mandDef} and the Levi-Civita contraction $\trFive$ 
is defined in \cref{eq:tr5}. We discuss the details of our choice of
master-integral basis in \cref{app:masterIntegralBasis}.
We then make the ansatz that
\begin{equation}\label{eq:MICoeffExpansion}
	c_{i}(\epsilon,\vec s,\trFive)
	=\frac{1}{P_i(\epsilon)}\sum_{k=0}^{\kappa_i}
	\epsilon^k\,c_{i,k}(\vec s,\trFive)\,,
\end{equation}
where we used the fact that there are no poles in $\epsilon$ that are kinematic
dependent, and $\kappa_i$ is the maximal power of $\epsilon$ in the numerator.
The polynomials $P_i(\epsilon)$ are trivial to determine by sampling the coefficients
at enough values of $\epsilon$ (for details see e.g.\ \cite{Abreu:2020xvt}).
Since $\trFive$ can be written as the square root of a polynomial in 
the Mandelstam variables, it follows that the most generic coefficient 
$c_{i,k}$ can be written as
\begin{equation}\label{eq:MICoeffExpansionParityDecomposition}
	c_{i,k}(\vec s,\trFive)
	=c_{i,k}^{+}(\vec s)+\trFive\,c_{i,k}^{-}(\vec s)\,,
\end{equation}
where $c_{i,k}^+(\vec s)$ and $c_{i,k}^-(\vec s)$ are rational functions of the $s_{ij}$.
We note that 
\begin{align}\begin{split}
  c_{i,k}^+(\vec{s})&=\frac{1}{2}\Big(c_{i,k}\left(\vec{s}, \trFive\right)+ c_{i,k}\left(\vec{s}, -\trFive\right)\Big),\\
  c_{i,k}^-(\vec{s})&=\frac{1}{2\,\mathrm{tr}_5}\Big(c_{i,k}\bigl(\vec{s},\trFive\bigr)-
  c_{i,k}\bigl(\vec{s}, -\trFive\bigr)\Big),
\end{split}\end{align}
which means that we can access individually each of these rational functions
of the $s_{ij}$ by evaluating the amplitudes at parity-conjugate phase-space points.

To determine the analytic form of the rational functions $c_{i,k}^\pm$, 
we numerically evaluate the master integral coefficients using the two-loop
numerical unitarity method outlined in \cref{sec:2lNumUni}.
We recall that these numerical evaluations are performed using finite-field arithmetic
which allows us to obtain exact values for the coefficients \cite{Peraro:2016wsq}.
We use the variables defined in appendix C of~\cite{Abreu:2019odu} to obtain
a rational parametrization of phase space. This parametrization presents 
the major advantage of having all but one twistor variable being 
Mandelstam invariants, which removes any ambiguity in mapping an expression from 
twistor variables to Mandelstam invariants.  Finally, in all
calculations we set $s_{12}=1$ and work with dimensionless variables. The
dependence on $s_{12}$ is reintroduced by dimensional analysis.

As in ref.~\cite{Abreu:2019odu}, we find that the denominator of the 
$c_{i,k}^\pm(\vec s)$ can be easily determined. Indeed, we find that
\begin{align}\label{eq:ripmAns}
  c_{i,k}^\pm(\vec{s})=\frac{n_{i,k}^\pm(\vec{s})}{\prod_j W^{q_{i,k,j}}_j(\vec{s})}\,,
\end{align}
where the $W(\vec s)$ are a subset of the so-called letters of the symbol alphabet
associated with the contributing master integrals, namely the subset that is polynomial
in $\vec s$ and not $\trFive$.\footnote{%
  Denominators of the form $\trFive^n$ with $n$ even are allowed since they are polynomials in $\vec{s}$,
  see also the discussion in \cref{sec:2lrem}.
}
As the photons are not
ordered, it is not sufficient to consider the planar alphabet which is only closed
under cyclic permutations. The full non-planar alphabet \cite{Chicherin:2017dob},
however, is closed under
all photon permutations and  sufficient to determine all the denominators. The
powers $q_{i,k,j}$ are determined by evaluating the amplitudes on a
one-dimensional line in phase-space \cite{Abreu:2018zmy,Abreu:2019odu}.

The determination of the analytic form of the coefficients
$c_{i,k}$ is then reduced to the determination of the polynomials
$n_{i,k}^\pm(\vec{s})$, which depend on four variables (once we set $s_{12}=1$).
With our in-house implementation of the multivariate Newton method \cite{Abreu:2020xvt},
we obtain their analytic form with coefficients in a finite field.
Following the simplification procedure of the expressions described in
ref.~\cite{Abreu:2019odu},
we were able to reconstruct all rational coefficients from a small number of numerical evaluations in additional finite fields.
We provide these results in ancillary files.

\begin{table}
  \centering
  \begin{tabular}{cccc}
    \toprule
    Helicity & Max degree & \# independent $c_{i,k}^\pm$ \\
    \midrule
    $A^{(2,0)}_{-++}$ & 32 & 1320  \\
    $A^{(2,\NF)}_{-++}$ & 20 & 203  \\[.5mm]
    $A^{(2,0)}_{+++}$ & 27 & 1244 \\[.5mm]
    $A^{(2,\NF)}_{+++}$ & 18 & 130  \\[.5mm]
    \bottomrule
  \end{tabular}
    \caption{Characterizing data for analytic expressions of master integrals, 
    see \cref{eq:Abis,eq:MICoeffExpansion}.
    `Max degree' is the highest polynomial degree across the different numerators 
   $c_{i,k}^\pm$. The column `\# independent $c_{i,k}^\pm$' gives the dimension of 
   the space of rational functions required to write all the $c_{i,k}^\pm$.
   }
    \label{tab:coeffsComplexity}
\end{table}

In \cref{tab:coeffsComplexity} we compile information that characterizes the
complexity of the master integral coefficients.
The column labeled `Max degree' is a measure of the 
complexity of the functional reconstruction step. The column 
`\# independent $c_{i,k}^\pm$' is a measure of the complexity of the final
result. We find that the complexity of the reconstruction is comparable
to the two-loop remainders of five-parton amplitudes 
reconstructed in ref.~\cite{Abreu:2019odu},
but the complexity of the final result is much higher than in the five-parton
remainder case (where the number of independent rational functions was at most
$\mathcal{O}(120)$).

\subsection{Two-loop Finite Remainders}
\label{sec:2lrem}

Once the coefficients $c_{i}$ in \cref{eq:Abis} have been determined,
and the expressions for the master integrals in terms of pentagon functions
are known, we can obtain the decomposition of the two-loop 
remainder in terms of pentagon functions, see \cref{eq:pentFunc}.
The coefficients $r_i$ have a form similar to the $c_{i,k}$, that is
\begin{equation}\label{eq:ripmDef}
	r_i(\vec s, \trFive)=r_i^+(\vec s)+\trFive\,r_i^-(\vec s)\,,
\end{equation}
where the $r_i^+(\vec s)$ and $r_i^-(\vec s)$ are rational functions of the $s_{ij}$.
As for the $c_{i,k}^\pm$, the denominator of the $r_{i}^\pm$ is given
by products of the letters in the subset of the (non-planar) symbol 
alphabet that are polynomial in 
the $s_{ij}$.
It is interesting to note that the denominator $\trFive^2$ is absent,
even though it is polynomial in the Mandelstam invariants. This is not the case
for the $c_{i,k}^\pm$, that is there are poles in the coefficients
$c_{i}(\epsilon,\vec s,\trFive)$ that are absent in the 
$r_{i}(\vec s,\trFive)$, and the $r_{i}$
are thus expected to be numerically more stable than the $c_{i}$.
We also note that, owing to the complexity of the rational functions in $c_{i,k}^\pm$,
it is a non-trivial task to obtain the expressions for the $r_i$ from them.
Finally, it is important to use the simplification procedure
described in ref.~\cite{Abreu:2019odu} to obtain relatively compact 
expressions suitable for efficient numerical evaluation.
Our expressions for the remainders are provided in ancillary files.

\begin{table}
  \centering
  \begin{tabular}{cccc}
    \toprule
    Helicity & Max degree & \# independent $r_i^\pm$ & Max weight \\
    \midrule
    $R^{(2,0)}_{-++}$ & 30 & 171 & 4 \\
    $R^{(2,\NF)}_{-++}$ & 13 & 57 & 3 \\[.5mm]
    $R^{(2,0)}_{+++}$ & 16 & 62 & 2 \\[.5mm]
    $R^{(2,\NF)}_{+++}$ & 12 & 12 & 1 \\[.5mm]
    \bottomrule
  \end{tabular}
    \caption{Characterizing data for analytic expressions of two-loop remainders,
    see \cref{eq:remaindersDef} and the discussion below for the notation.
   We use the same labels as in \cref{tab:coeffsComplexity}.
   `Max weight' is the highest transcendental weight
    of the pentagon functions appearing in each remainder.
   }
    \label{tab:remainderComplexity}
\end{table}

In \cref{tab:remainderComplexity} we compile some data to characterize the
complexity of the analytic expressions.
Compared to \cref{tab:coeffsComplexity}, we added the column `Max weight'
which gives a measure of the complexity of the contributing pentagon 
functions.\footnote{
   Roughly speaking, the transcendental weight corresponds to the number
   of iterated integrations in the definition of the pentagon functions.
 }
The highest possible transcendental weight in a two-loop remainder is 4, and only
one remainder saturates this bound. This number is also important
for numerical evaluations, as the highest 
weights dominate the evaluation time of the pentagon functions. 
It is interesting to note
that the maximal polynomial degree of the numerator we need to reconstruct
is only slightly lower than in the case of master integral coefficients, 
see \cref{tab:coeffsComplexity},
which means that the complexity of the reconstruction of remainders and of 
master integral  coefficients is not substantially different.
The complexity of the final result for master integrals 
is however much higher, since they depend on a larger number of independent rational 
functions. As expected, for NNLO
applications it is thus much more efficient to work at the level of the remainders.

\subsection{Reference Values}\label{sec:refVal}
In order to facilitate the comparison with our results and to explicitly
demonstrate the pole structure of the amplitudes, we present a
numerical evaluation of the remainders and loop amplitudes on a randomly-chosen
phase-space point in the physical region. 
Given that we are computing the Lorentz-invariant quantities
defined in \cref{eq:leadingColourDec}, 
it is sufficient to specify the five independent Mandelstam variables
together with the value of $\tr_5$,
\begin{align}
  \begin{split}
    s_{12} &=  1.322500000, \quad \,\,
    s_{23} =  -0.994109498, \quad \,\,
    s_{34} =   0.264471591, \\
    s_{45} &=  0.267126049, \quad \,\,
    s_{15} =  -0.883795230, \quad \,\,
    \tr_5  =  -0.11382836\ii.
  \end{split}
  \label{eq:referencePoint}
\end{align}
The one- and two-loop amplitudes and remainders
evaluated on this point are presented in \cref{tab:referenceValuesAmp1Loop,tab:referenceValuesAmp2Loop,tab:referenceValuesRemainders}.
Expressions for the amplitudes in terms of master integrals for the one- and
two-loop amplitudes are presented in a series of ancillary files in the
directories \texttt{anc/oneLoopAmplitudes/} and \texttt{anc/twoLoopAmplitudes/}
respectively. Expressions for the the finite remainders in terms of pentagon
functions are presented in a series of files in the directory
\texttt{anc/remainders/}. We show how to assemble these files into the full
amplitudes and remainders in \texttt{anc/example\_assembly.m}, where, 
using an included numerical evaluation of the master integrals and 
pentagon functions, the expressions are combined to compute the amplitudes 
and remainders at the
reference phase-space point \eqref{eq:referencePoint} and reproduce the numbers in
\cref{tab:referenceValuesAmp1Loop,tab:referenceValuesAmp2Loop,tab:referenceValuesRemainders}.

\begin{table}[tbp]
  \centering
  \small
  \newcolumntype{L}{>{\centering\arraybackslash}m{15ex}<{}}
  \renewcommand{\arraystretch}{1.3}
  \begin{tabular}{lLLLLL}
    \toprule
  & $\epsilon^{-2}$ & $\epsilon^{-1}$ &  $\epsilon^0$ &  $\epsilon^{1}$ &  $\epsilon^{2}$ \\
  \midrule
    $A^{(1)}_{-++}$ & $-1.000000000$ &  $-3.174284697$ $-3.141592654\ii$ &  $-3.437681197$ $-16.69077768\ii$ &  $-4.542364174$ $-48.29215997\ii$ &  $-28.34154945$ $-104.73071151\ii$ \\
    $A^{(1)}_{+++}$ &  $0$ &  $0$ &  $-122.4876141$ $-218.2099911\ii$ &  $-613.1620024$ $-1772.249665\ii$ &  $-1264.781477$ $-6727.583766\ii$ \\
    \bottomrule
  \end{tabular}
  \caption{Reference evaluations of all independent bare one-loop amplitudes on the phase-space point of \cref{eq:referencePoint}.}
  \label{tab:referenceValuesAmp1Loop}
\end{table}

\begin{table}[tbp]
  \centering
  \small
  \newcolumntype{L}{>{\centering\arraybackslash}m{15ex}<{}}
  \renewcommand{\arraystretch}{1.3}
  \begin{tabular}{lLLLLL}
    \toprule
      &  $\epsilon^{-4}$ &  $\epsilon^{-3}$ &  $\epsilon^{-2}$ &  $\epsilon^{-1}$ &  $\epsilon^{0}$ \\
    \midrule
    $A^{(2,0)}_{-++}$ &  $0.500000000$ &  $2.257618031$ $+ 3.141592654\ii$ & $-3.317245357$ $+ 20.90350063\ii$ &  $-55.54942686$ $+ 44.34772277\ii$ & $-248.7699347$ $-87.79211669\ii$ \\
    $A^{(2,\NF)}_{-++}$ &  $0$ &  $0.1666666667$ &  $1.335872677$ $+1.047197551\ii$ &  $4.646264515$ $+12.872514370 \ii$ & $10.33373680$ $+83.15472523 \ii$ \\
    $A^{(2,0)}_{+++}$ &  $0$ &  $0$ &  $122.4876141$ $+218.2099911\ii$ &  $-132.6755953$ $+2049.613188\ii$ &  $-9927.845724$ $+3575.607623\ii$ \\
    $A^{(2,N_f)}_{+++}$ &  $0$ &  $0$ &  $0$ &  $81.65840942$ $+145.4733274\ii$&  $895.9475013$ $+2327.538099\ii$ \\
    \bottomrule
  \end{tabular}
  \caption{Reference evaluations of all independent bare two-loop amplitudes on the phase-space point of \cref{eq:referencePoint}.}
  \label{tab:referenceValuesAmp2Loop}
\end{table}

\begin{table}[tbp]
  \centering
  \renewcommand{\arraystretch}{1.3}
  \begin{tabular}{>{\centering\arraybackslash}m{10ex}>{\centering\arraybackslash}m{40ex}}
    \toprule
    $R^{(1)}_{-++}$ & $-5.281908761-6.718468192\ii$ \\
    $R^{(1)}_{+++}$ & $-122.4876141-218.2099911\ii$ \\
    \midrule
    $R^{(2,0)}_{-++}$ & $-17.93042514-84.48074943\ii$ \\
    $R^{(2,\NF)}_{-++}$ & $8.536235118+25.51694192\ii$ \\
    $R^{(2,0)}_{+++}$ & $-2043.581205-3461.464426\ii$ \\
    $R^{(2,N_f)}_{+++}$ & $327.6279319+861.8112864\ii$ \\
    \bottomrule
  \end{tabular}
  \caption{Reference evaluations of all independent one-loop and two-loop remainders on the phase-space point of \cref{eq:referencePoint}.}
  \label{tab:referenceValuesRemainders}
\end{table}

\subsection{Validation}

In this section we discuss the checks we performed on our setup to compute
the two-loop amplitudes for three-photon production at hadron colliders.

We first computed analytic four- and five-point one-loop amplitudes. 
The four-point amplitudes
were checked against the results of ref.~\cite{Glover:2003cm}, 
and we found full agreement with the quoted remainders.
For the five-point amplitudes, we reproduced the results obtained from 
\texttt{OpenLoops} \cite{Buccioni:2019sur} to order $\epsilon^0$ and numerically verified all the 
relations in \cref{tab:hel_rel}.
Finally, we checked that five-point one-loop amplitudes have the correct
collinear behavior up to order $\epsilon^2$, i.e., that in these limits they
are either regular or factorize into products of splitting amplitudes and
four-point amplitudes.

At the two-loop level, we first recomputed the four-point amplitudes, reproducing
the two-loop remainders given in ref.~\cite{Glover:2003cm}. 
For the five-point amplitudes, we verified that they have the correct
pole structure and numerically checked that they satisfy the relations 
in \cref{tab:hel_rel}. 
To further check the results at order $\epsilon^0$, we also verified
that the amplitudes have the correct collinear behavior as described 
in \cref{sec:collChecks}.
Finally, since ref.~\cite{Chawdhry:2019bji}
does not include explicit results for the two-loop amplitudes our expressions cannot
be directly compared to theirs. Nevertheless,  
our results were shown to lead to consistent predictions for the production 
of three photons at the LHC \cite{Kallweit:2020gcp}.



\section{Numerical Evaluation}\label{sec:results}

\paragraph{Squared Finite Remainders}

The contribution of two-loop helicity amplitudes to physical cross sections is constructed
from the finite remainders (see e.g.\ \cite{Weinzierl:2011uz}) -- specifically
the squared finite remainders, summed over helicity 
and color states. This object, which we denote $H$,
admits a perturbative expansion in powers of the renormalized coupling~$\alpha_s$,
\begin{equation}
  H=H^{(0)}+\frac{\alpha_s}{2\pi}H^{(1)}+\left(\frac{\alpha_s}{2\pi}\right)^2H^{(2)}
  +\mathcal{O}(\alpha_s^3)\,,
\end{equation}
which we normalize such that $H^{(0)}=1$. The $\mathcal{O}(\alpha_s)$ 
contribution $H^{(1)}$ is given by
\begin{equation}\label{eq:h1}
  H^{(1)} = \frac{1}{\sum_h \abs{\mathcal{\mathcal{M}}_h^{(0)}}^2} \, 
  C_F \left(\sum_h \abs{\mathcal{M}^{(0)}_h}^2 ~ 2 \Re\left[R^{(1)}_h\right]\right)\,,
\end{equation}
and the $\mathcal{O}(\alpha_s^2)$ contribution $H^{(2)}$ by
\begin{align}\begin{split}\label{eq:h2}
  H^{(2)} = \frac{1}{\sum_h \abs{\mathcal{M}_h^{(0)}}^2} 
    &\left(\sum_h \frac{\NC^2}{4} \NC \abs{ \Phi_h R_h^{(1)}}^2 \right.+ \\
    &\left.\sum_h\abs{\mathcal{M}^{(0)}_h}^2 ~ 2 \Re\left[\frac{N_c^2}{4}  R^{(2,0)}_h +  C_F T_F \NF R_h^{(2,\NF)}\right] \right)\,.
\end{split}\end{align}
In this expression, the first line corresponds to the one-loop squared
contribution, and the second line to the interference of two-loop and
tree-level amplitudes. In the one-loop squared contributions, we have expanded
the $C_F^2$ factor for consistency with the limit in which the two-loop
contributions were computed.
The normalization factor in \cref{eq:h1,eq:h2}
is consistent with setting $H^{(0)}=1$, and in our normalization
\begin{equation}
  \abs{\mathcal{M}_h^{(0)}}^2=e_q^6\NC\abs{\Phi_h}^2\abs{A^{(0)}_h}^2\,.
\end{equation}
The helicity sums in \cref{eq:h1,eq:h2} are performed using the relations from \cref{tab:hel_rel}.
We note that the remainders $R^{(2,0)}_{+++}$ and $R^{(2,\NF)}_{+++}$
do not contribute to $H^{(2)}$, in the same way
that $R^{(1)}_{+++}$ does not contribute to $H^{(1)}$. Nevertheless,
$R^{(1)}_{+++}$ does contribute to $H^{(2)}$ through the one-loop squared
contributions.

\paragraph{Numerical Evaluation}

Having phenomenological applications in mind, we have 
implemented in a \texttt{C++} library the numerical evaluation of the finite remainders 
$R^{(1,0)}_h$, $R^{(2,0)}_h$, $R^{(2,\NF)}_h$ (see \cref{eq:remaindersDef}),
and of $H^{(1)}$ and $H^{(2)}$ as defined in \cref{eq:h1,eq:h2}.
This library can be obtained from a \texttt{git} repository \cite{FivePointAmplitudes}. 
The library relies on \texttt{PentagonFunctions++}~\cite{Chicherin:2020oor} 
for the numerical evaluation of pentagon functions.
For installation and usage instructions we refer to the \verb|README.md| file which can be found in the root directory of the repository. 

We recall that the rational coefficients in the decomposition
of the remainders in terms of pentagon functions are simplified using 
the multivariate partial fractioning procedure outlined in \cite{Abreu:2019odu}.
Furthermore, we optimize the evaluation of rational coefficients using 
\texttt{FORM}~\cite{Ruijl:2017dtg,Kuipers:2013pba}.
As a result, the time spent on their numerical evaluation is 
negligible compared to the time spent on the evaluation of transcendental functions.
On average, the evaluation time of $H^{(2)}$ is on the order of a few seconds 
per phase-space point in double precision, using the default 
settings of \texttt{PentagonFunctions++}.

\begin{figure}[ht]
  \centering
  \includegraphics[width=0.8\textwidth]{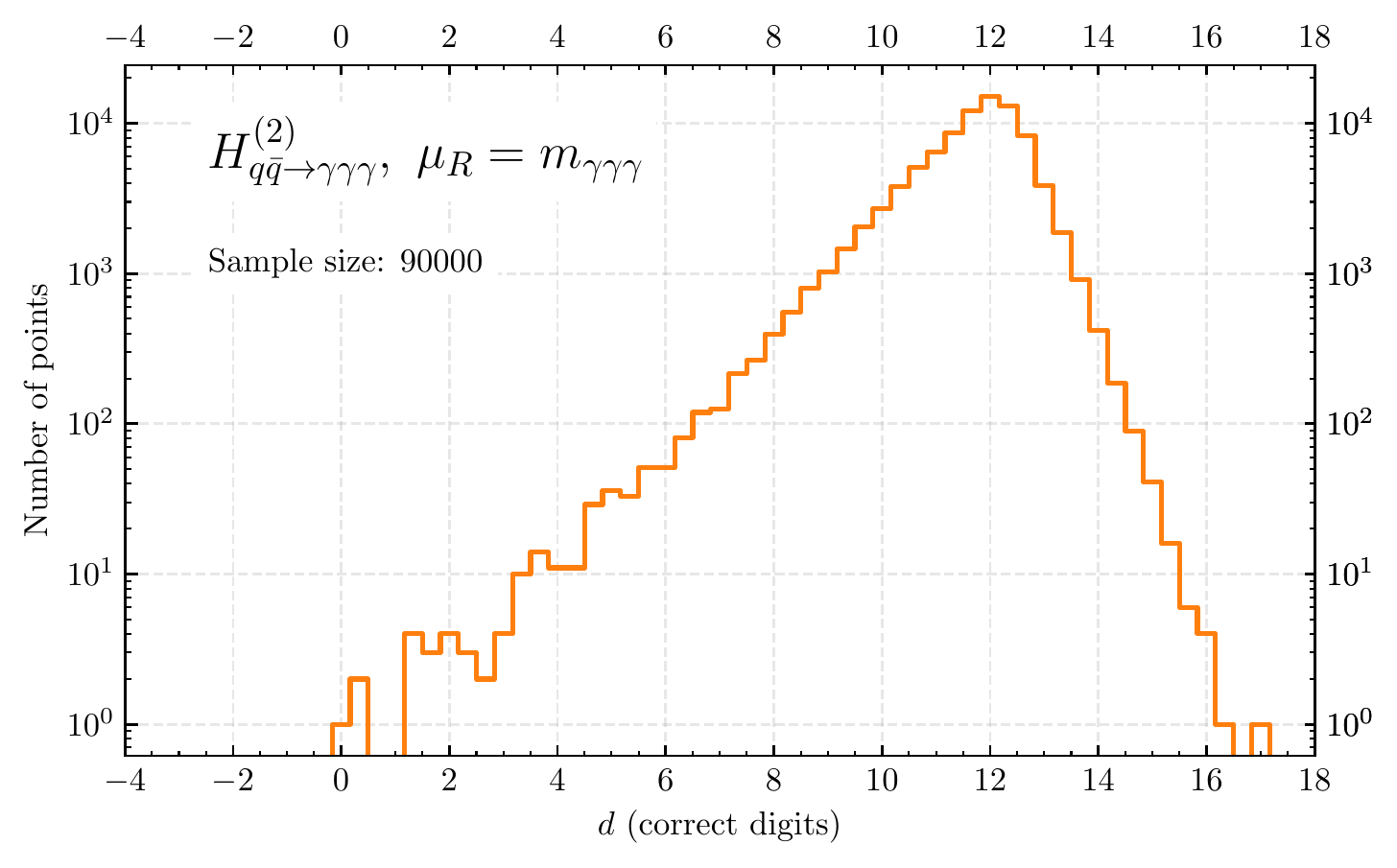}
  \caption{
    The logarithmic distribution of correct decimal digits (as defined in \cref{eq:relerr}) for 90000 double-precision evaluations of the $H^{(2)}$ function.
    The phase-space points are sampled from a distribution representative of typical phenomenological studies.
  }
  \label{fig:H2stability}
\end{figure}

Besides the evaluation speed, the calculation of the $H^{(i)}$ must be numerically 
stable. To demonstrate the stability of our results, we compare
the numerical evaluation of $H^{(2)}$ in double precision,
which we denote $H^{(2)}_\text{double}$, with the evaluation 
in quadruple precision, which we denote $H^{(2)}_\text{quad}$, on a sample
of 90000 phase-space points from the distribution employed in \cite{Kallweit:2020gcp} for the computation of predictions at the center-of-mass energy of 8 TeV.
The phase-space cuts correspond to the ones used in the ATLAS 8 TeV measurement \cite{Aaboud:2017lxm} (we refer to the table 1 of \cite{Kallweit:2020gcp} for the explicit definitions).
We set $\NC=3$, $\NF=5$, and the renormalization scale $\mu_R$ to the 
invariant mass of the three-photon system $m_{\gamma\gamma\gamma}$.
Assuming $H^{(2)}_\text{quad}$ to be correct at least up to a relative error of $\sim 10^{-16}$, we define
\begin{equation}\label{eq:relerr}
  d = -\log_{10}\abs{ \frac{H^{(2)}_\text{double}}{H^{(2)}_\text{quad} } - 1 }
\end{equation}
as a measure of the number of correct decimal digits in $H^{(2)}_\text{double}$. 
In \cref{fig:H2stability} we show a histogram of this quantity for the 
90000 sampled points on a logarithmic scale.
We observe that less than $0.1\%$ of the sampled points have an accuracy 
of less than four digits. 
This level of accuracy is more than adequate for phenomenological applications.
Indeed, our implementation was already used for a Monte Carlo phase-space integration 
in \cite{Kallweit:2020gcp}, which converged to an overall integration error 
below $1\%$ in NNLO differential distributions.

It is interesting to note that a good understanding of the physical properties
that govern the analytic structure of scattering amplitudes can be used
to explain and improve the behavior of numerical algorithms. In particular, we recall
that our analytic representation of the remainders (and thus of the $H^{(i)}$)
only has poles that are associated with a subset of the letters of
the two-loop five-point massless alphabet \cite{Chicherin:2017dob}, 
see the discussion in \cref{sec:2lrem}. 
While the letters corresponding to unphysical or spurious singularities can vanish 
inside the physical phase space, the amplitude must stay regular
As an example, $s_{12}+s_{23}$ is a letter which can vanish in the physical
phase space, see \cref{eq:physregion}, but is not a physical threshold.
This implies that large cancellations can potentially occur  
when the amplitude is evaluated on phase-space points that
are close to the surfaces where those letters vanish. Conversely,
away from these small neighborhoods, the numerical evaluations
should be accurate. We observe precisely this behavior.
We verified that all of the unstable phase-space points 
in \cref{fig:H2stability} are close 
to surfaces associated with spurious singularities.
This shows that it is important to organize the analytic structure of the 
rational coefficients
in the expressions for amplitudes or remainders using physical considerations,
as it allows us to obtain compact expressions whose numerical evaluation is under
full control. Indeed,
once such a form is found, we can setup a robust precision-rescue system
based on the fact that all the problematic regions of phase-space are explicitly known.
We leave this for future work.


\section{Conclusion}\label{sec:conclusion}

In this paper we have computed the two-loop planar corrections to the production
of three photons at hadron colliders. 
This was achieved within the framework of two-loop numerical unitarity, coupled
with analytic reconstruction techniques. Our results include 
expressions for both one- and two-loop amplitudes valid to all orders in the  dimensional 
regulator. In both cases, the amplitudes are written in terms of a set of master
integrals. To our knowledge, this is the first time that master-integral coefficients
have been obtained in analytic form for physical two-loop five-point scattering 
amplitudes. All our results are presented in a set of ancillary files.

As is well known, for NNLO (two-loop) phenomenology only the finite remainders of two-loop amplitudes are needed.
By writing one- and two-loop amplitudes in terms of a basis of 
pentagon functions up to the required order in $\epsilon$,
we obtained a decomposition of the remainder in terms of these functions.
After simplifying the coefficients in this decomposition using multivariate 
partial-fractioning techniques,
we obtained compact expressions for the two-loop
remainders of the two independent helicity amplitudes.
We demonstrated that the remainders are simpler functions than the all-order amplitudes.

While over the last years there has been a number of new results for 
massless five-point amplitudes, when having in mind phenomenological applications
it is also important that the expressions  for the remainders are numerically 
stable and can be efficiently evaluated across the relevant physical phase space.
This point has been a major obstacle to computing the NNLO corrections for those
processes.
The expressions we obtain in this paper are ready to be used for phenomenological
studies. This  was demonstrated by verifying the numerical stability of the remainders
when combined in a code that computes the color- and helicity-summed squared 
remainders.
We have made this code public in a format that can be interfaced with
real-radiation programs and employed to compute complete NNLO theoretical predictions.
This is the first time that a two-loop massless five-point process
is available in this form. Our results
have already been used in ref.\ \cite{Kallweit:2020gcp}, and we expect
that they will be instrumental in any future studies.


\section*{Acknowledgments}

We thank Fernando Febres Cordero for many enlightening 
discussions and assistance in setting up the
cluster runs, and we thank Harald Ita for many discussions and 
comments on this manuscript.
The work of B.P~is supported by the French Agence Nationale pour la Recherche, 
under grant ANR–17–CE31–0001–01.
V.S.~is supported by the European Research Council (ERC) under the European 
Union's Horizon 2020 research and innovation programme,
\textit{Novel structures in scattering amplitudes} (grant agreement No.\ 725110).
The authors acknowledge support by the state of Baden-Württemberg through bwHPC
and the German Research Foundation (DFG) through grant no INST 39/963-1 FUGG (bwForCluster NEMO).

\appendix

\section{Master Integral Basis}
\label{app:masterIntegralBasis}
\FloatBarrier

Here we give details on the bases of master integrals employed in the ancillary files.
We consider a master integral decomposition at both one and two loops of the
form in  \cref{eq:A}, where the master integrals within the basis are labelled
$m_{\Gamma,i}$.
Each propagator structure $\Gamma$ can be understood diagrammatically, and they
can be grouped into sets where the elements are related by 
different permutations of the external legs.
We call each such grouping a topology.
We recall that the index $i$ in $m_{\Gamma,i}$ labels the different
master integrals associated with a given $\Gamma$.
In the ancillary files, master integrals are labelled by a topology name,
the index $i$, and the relevant scattering kinematics in terms of
five-point Mandelstam invariants. This information appears in the format
\begin{center}
  \texttt{MI[topologyName, i, mandelstam1, mandelstam2, \ldots]}.
\end{center}
In this representation, the permutations act on the Mandelstam
arguments and hence permutations of master integrals are
codified as different Mandelstam invariants appearing in the final arguments.

Each $m_{\Gamma,i}$ denotes a numerator.
These are polynomial in the
loop momenta, and rational in external kinematic variables
and in the dimensional regulator $D$. 
The notation used in specifying the numerators in the ancillary files is defined in 
table \ref{tab:numeratorObjects}.
There, we make use of the scalar product of the components of the loop momenta beyond
4-dimensions
\begin{equation}
  \mu_{ij} = \ell_i^{D-4} \cdot \ell_j^{D-4} \, .
\end{equation}
We collect the definitions of the master integrals employed in the one- and 
two-loop amplitudes in the ancillary files
\texttt{anc/oneLoopAmplitudes/MasterIntegralDefinitions.m} and
\texttt{anc/twoLoopAmplitudes/MasterIntegralDefinitions.m}, in the format
\begin{center}
  \texttt{MI[\ldots] -> num},
\end{center}
where $\texttt{num}$ is the expression of the numerator.
All master integrals are ``unitarity compatible'', in that they have unit
propagator powers, and we specify the numerator $m_{\Gamma,i}(\ell)$ for a
single representative of each topology.
We present the external-momenta conventions 
and loop-momenta routing for a representative
permutation of the one-loop integrals in \cref{tab:oneLoopMasterRoutings},
and of the two-loop integrals in  
\cref{tab:masterRoutings5,tab:masterRoutings4,tab:masterRoutings23}. In
all cases, we take all external momenta as outgoing and we omit the loop
momentum labels when they are not required to disambiguate the numerators.

\begin{table}[htbp]
\renewcommand{\arraystretch}{1.4}
\centering
\begin{tabular}{c  c}
\texttt{l[i]} & $\ell_i$ \\
  \hline
\texttt{p[k]} & $p_k$ \\
  \hline
\texttt{sq[v]} & $v^2$ \\
  \hline
\texttt{sp[v[i], v[j]]} & $v_i \cdot v_j$ \\
  \hline
\texttt{tr5[1, 2, 3, 4]} & $\tr_5$ \\
  \hline
\texttt{mu[i, j]} & $\mu_{ij}$ \\
  \hline
\texttt{D} & $D$ \\
\end{tabular}
\caption{Notation used in specifying the numerators of master integrals.}
\label{tab:numeratorObjects}
\end{table}

\begin{table}[htbp]
\centering
{
\renewcommand{\arraystretch}{3}
\begin{tabular}{c | c}
\texttt{MI["Bubble", i, s12]} & $\eqnDiag{\includegraphics[scale=0.5]{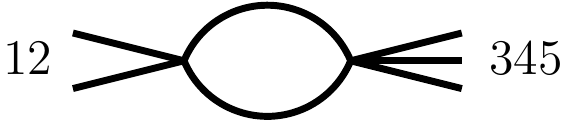}}$ \\
\texttt{MI["OneMassTriangle", i, s12]} & $\eqnDiag{\includegraphics[scale=0.5]{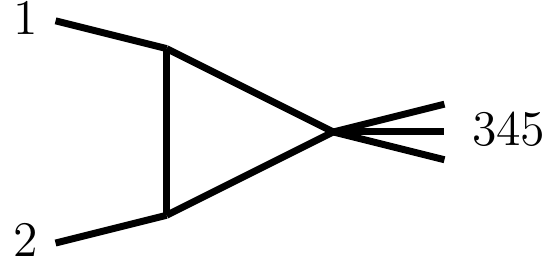}}$ \\
\texttt{MI["TwoMassTriangle", i, s12, s45]} & $\eqnDiag{\includegraphics[scale=0.5]{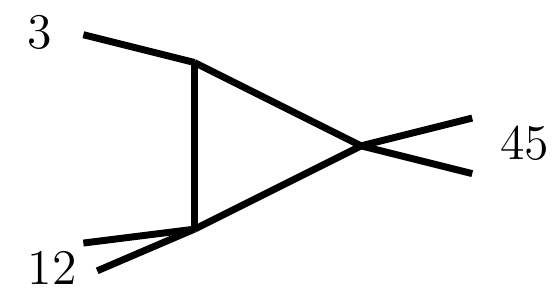}}$ \\
\texttt{MI["OneMassBox", i, s45, s12, s23]} & $\eqnDiag{\includegraphics[scale=0.5]{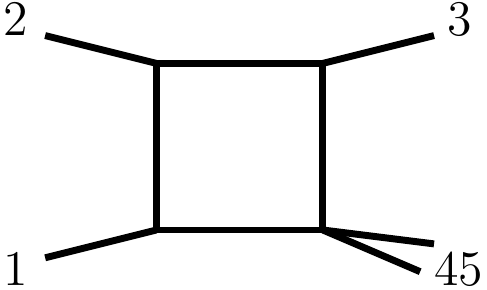}}$ \\
\texttt{MI["Pentagon", i, s12, s23, s45, s45, s15]} & $\eqnDiag{\includegraphics[scale=0.5]{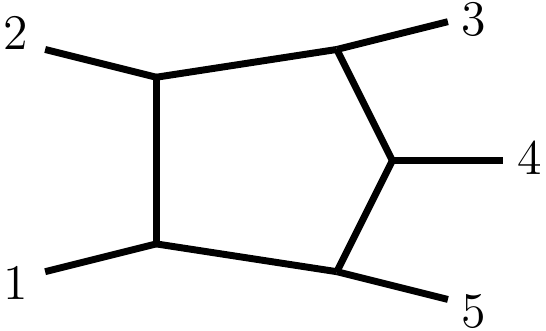}}$ \\
\end{tabular}
}
\caption{Permutation representatives for one-loop master integral topologies.
For one-loop master integrals we use the basis of \cite{Giele:2008ve},
which includes only a certain subset of IBP relations (those that are 
independent of $D$). The basis thus includes integrals that would be
reducible if one were to use the full set of IBP relations, such
as all \texttt{"OneMassTriangle"} masters. 
The advantage of using this basis is that
the coefficients in \eqref{eq:A} are $D$-independent.}
\label{tab:oneLoopMasterRoutings}
\end{table}

\begin{table}[htbp]
\centering
{
\renewcommand{\arraystretch}{3}
\begin{tabular}{c | c}
\texttt{MI["PentagonBox", i, s12, s23, s34, s15, s45]} & $\eqnDiag{\includegraphics[scale=0.5]{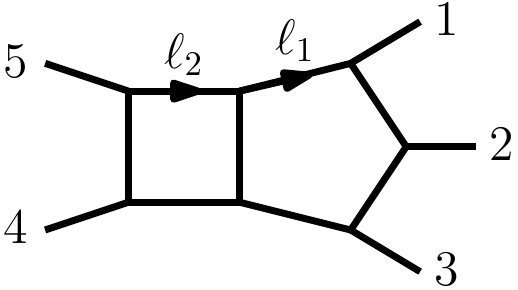}}$ \\
\texttt{MI["FivePointDoubleBox", i, s12, s23, s34, s15, s45]} & $\eqnDiag{\includegraphics[scale=0.5]{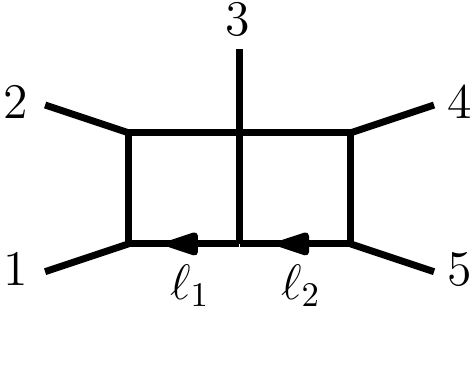}}$ \\
\texttt{MI["FivePointBoxTriangle", i, s12, s23, s34, s15, s45]} & $\eqnDiag{\includegraphics[scale=0.5]{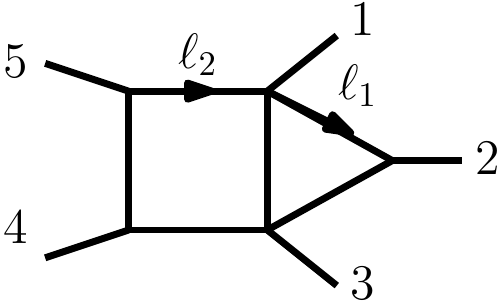}}$ \\
\texttt{MI["PentagonBubble", i, s12, s23, s34, s15, s45]} & $\eqnDiag{\includegraphics[scale=0.5]{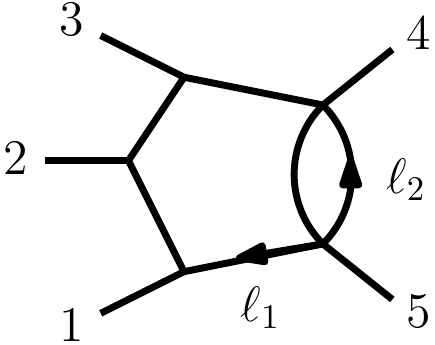}}$ \\
\end{tabular}
}
\caption{Permutation representatives for five-point two-loop master integral topologies}
\label{tab:masterRoutings5}
\end{table}

\begin{table}[htbp]
\centering
{
\renewcommand{\arraystretch}{3}
\begin{tabular}{c | c}
\texttt{MI["OneMassDoubleBox", i, s12, s34, s45]} & $\eqnDiag{\includegraphics[scale=0.5]{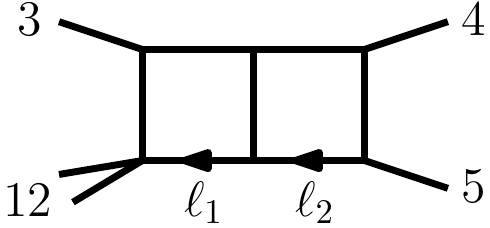}}$ \\
\texttt{MI["FourPointBoxTriangle", i, {s23, s14, s45}]} & $\eqnDiag{\includegraphics[scale=0.5]{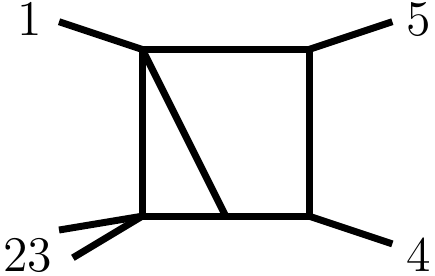}}$ \\
\texttt{MI["OneMassSlashedBoxHard", i, s45, s12, s23]} & $\eqnDiag{\includegraphics[scale=0.5]{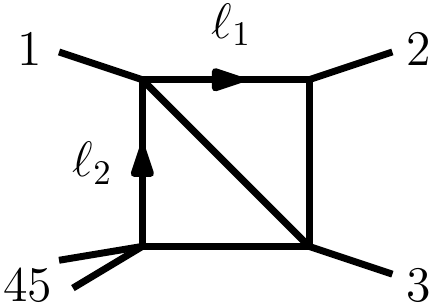}}$ \\
\texttt{MI["OneMassSlashedBoxEasy", i, s12, s34, s45]} & $\eqnDiag{\includegraphics[scale=0.5]{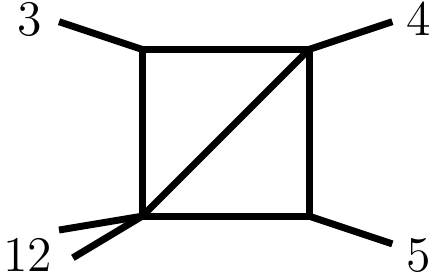}}$ \\
\texttt{MI["OneMassBoxBubbleHard", i, s23, s14, s45]} & $\eqnDiag{\includegraphics[scale=0.5]{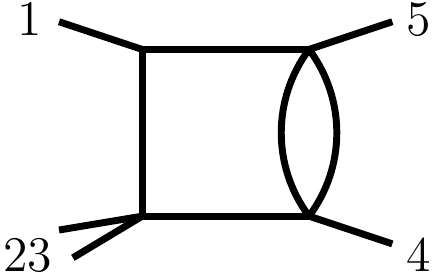}}$ \\
\texttt{MI["OneMassBoxBubbleEasy", i, s12, s34, s45]} & $\eqnDiag{\includegraphics[scale=0.5]{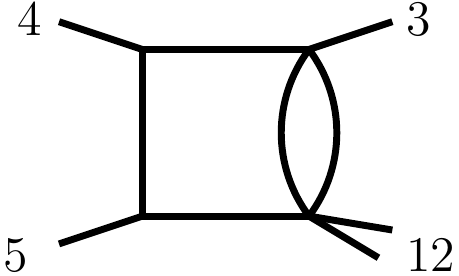}}$ \\
\texttt{MI["FactorizedBoxBubble", i, s45, s12, s23]} & $\eqnDiag{\includegraphics[scale=0.5]{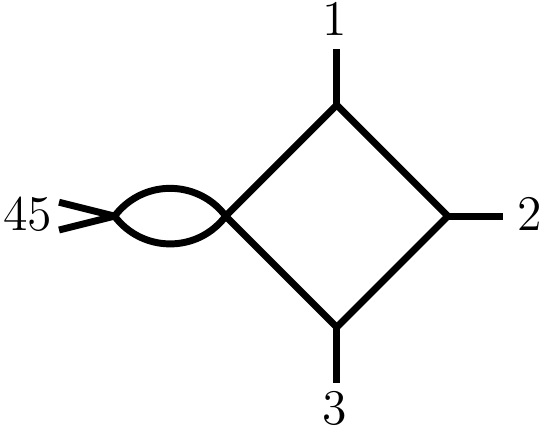}}$ \\
\end{tabular}
}
\caption{Permutation representatives for four-point two-loop master integral topologies.}
\label{tab:masterRoutings4}
\end{table}

\begin{table}[htbp]
\centering
{
\renewcommand{\arraystretch}{3}
\begin{tabular}{c | c}
\texttt{MI["TwoMassSlashedTriangle", 1, 4, s23, s45]} & $\eqnDiag{\includegraphics[scale=0.5]{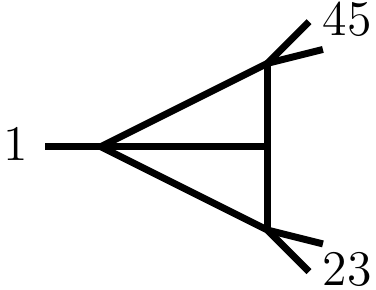}}$ \\
\texttt{MI["OneMassTriangleBubble", i, s45]} & $\eqnDiag{\includegraphics[scale=0.5]{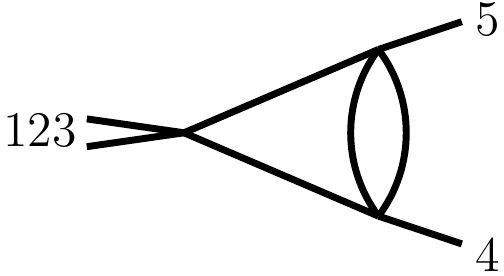}}$ \\
\texttt{MI["TwoMassTriangleBubble", i, s12, s45]} & $\eqnDiag{\includegraphics[scale=0.5]{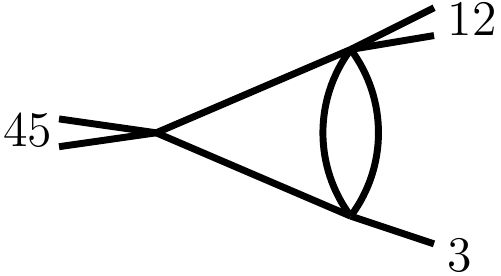}}$ \\
\texttt{MI["ThreePointFactorizedBubbleBubble", i, s23, s45]} & $\eqnDiag{\includegraphics[scale=0.5]{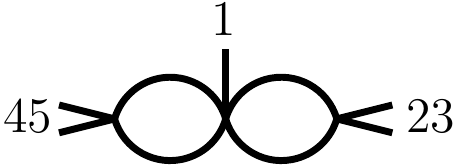}}$ \\
\texttt{MI["FactorizedBubbleBubble", i, s45]} & $\eqnDiag{\includegraphics[scale=0.5]{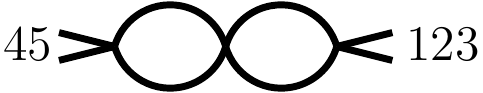}}$ \\
\texttt{MI["Sunrise", i, s34]} & $\eqnDiag{\includegraphics[scale=0.5]{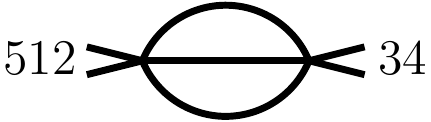}}$ \\
\end{tabular}
}
\caption{Permutation representatives for two- and three-point two-loop master integral topologies.}
\label{tab:masterRoutings23}
\end{table}

\FloatBarrier


\section{Collinear Checks}\label{sec:collChecks}

The behavior of scattering amplitudes with massless external legs in the limit
where two external legs become collinear is well understood.
In this appendix we discuss how this was used to check
our results for the amplitudes for the production of three photons
at hadron colliders.

Let us first review the universal behavior of scattering amplitudes
when two massless external legs with four-momenta $p_i$ and $p_j$ become collinear.
We denote this limit as $i||j$.
As they are massless, in the collinear limit the momenta become proportional.
We introduce the massless four-momentum $p_P$, such that when $i||j$ 
\begin{equation}\label{eq:colLimZ}
  p_i \rightarrow z p_P \quad \quad \mathrm{and} \quad \quad p_j \rightarrow (1-z) p_P\,.
\end{equation}
The variable $z$ can be understood as the fraction of energy of $p_P$ which is
contributed by~$p_i$.
We note that $z\rightarrow 0$ or 1 corresponds, respectively, to the limit where
particles $i$ or $j$ go soft. 

The study of the collinear behavior of bare helicity amplitudes is better formulated 
at  the level of color-stripped amplitudes. In our case, we define
\begin{equation}
	\bar{\mathcal{A}}(1_q^{h_1},2_{\bar{q}}^{h_2},3_\gamma^{h_3},4_\gamma^{h_4},5_\gamma^{h_5}):=
	\Phi(1_q^{h_1},2_{\bar{q}}^{h_2},3_\gamma^{h_3},4_\gamma^{h_4},5_\gamma^{h_5})
  \mathcal{A}(1_q^{h_1},2_{\bar{q}}^{h_2},3_\gamma^{h_3},4_\gamma^{h_4},5_\gamma^{h_5})
  \,,
\end{equation}
which is trivially related to the $\mathcal{M}$ and $\mathcal{A}$ we have used 
elsewhere in the paper, see \cref{eq:def_A}. In particular, we will use the same 
abbreviated notational style whenever it is unambiguous.
In the $i||j$ limit, the amplitudes can diverge.
The divergent behavior is captured by the factorization formula
\begin{equation}
\bar{\mathcal{A}}(i^{h_i}, j^{h_j}, \ldots) \xrightarrow{i || j} 
\sum_h \mathrm{Split}_{P^h}(i^{h_i}, j^{h_j}) 
\bar{\mathcal{A}}(\bar{P}^{-h}, \ldots)\,.
\label{eq:AllOrdersSplitting}
\end{equation}
Here, the factors $\mathrm{Split}_{P^h}(i^{h_i}, j^{h_j})$ are known as 
the splitting amplitudes for the particle $P$ with heliciy $h$ 
to split into particles $i$ with heliciy $h_i$ and $j$ with helicity $h_j$.
The splitting functions depend only on the momenta, helicity and color quantum 
numbers of particles $i$, $j$ and $P$, and the dependence on $p_P$ is solely in terms 
of the energy fraction $z$. If the amplitude on the left-hand side has $n$ external legs,
the amplitude on the right-hand side is a $(n-1)$-point amplitude, where
$\bar{P}$ denotes the conjugate particle to the particle $P$ in the subscript of the
splitting function.
Finally, we note that it might be that
the right-hand side of \cref{eq:AllOrdersSplitting} is zero, in which case
the amplitude on the left-hand side is regular in the $i||j$ limit.

\begin{figure}
  \begin{equation*}
    \eqnDiag{\includegraphics[scale=0.43]{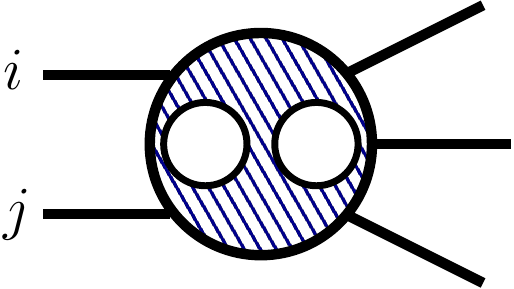}} \xrightarrow{i || j} \eqnDiag{\includegraphics[scale=0.43]{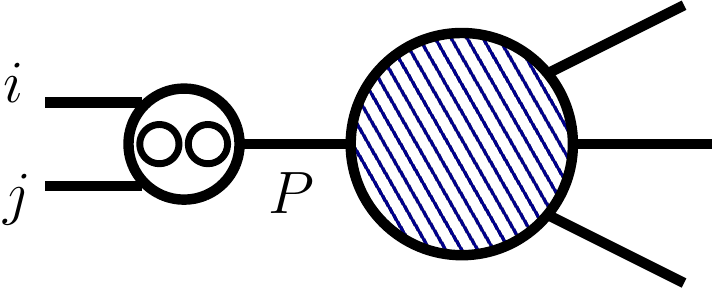}} + \eqnDiag{\includegraphics[scale=0.43]{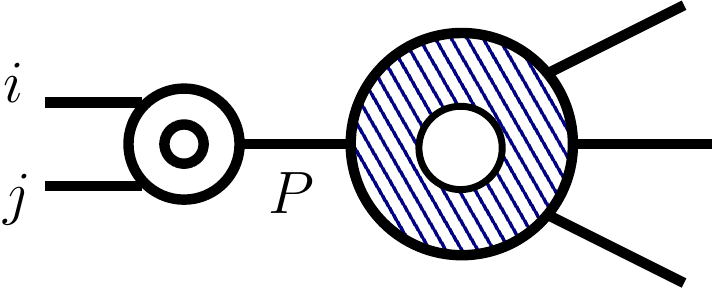}} + \eqnDiag{\includegraphics[scale=0.43]{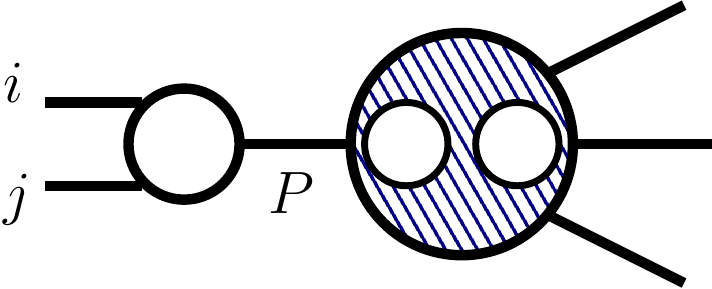}}
  \end{equation*}
  \caption{Graphical representation of equation \eqref{eq:twoLoopSplitting}.
White blobs represent the splitting amplitudes and hatched blobs represent
the four-point amplitudes. The number of contained circles represents the loop
order of the object. The types of particles are omitted from this representation,
the external and internal lines representing only momenta.}
  \label{fig:twoLoopSplitting}
\end{figure}

The behavior in \cref{eq:AllOrdersSplitting} is valid to all orders 
in the coupling constant. Naturally, the splitting amplitudes have a 
perturbative expansion in powers of $\alpha_s^0$, 
\begin{align}\begin{split}
\mathrm{Split}_{P^h}(i^{h_i}, j^{h_j}) = 
\mathrm{Split}^{(0)}_{P^h}(i^{h_i}, j^{h_j}) \,+\, 
&\,\frac{\alpha_s^0}{2 \pi}\,\mathrm{Split}^{(1)}_{P^h}(i^{h_i}, j^{h_j})\,+ \\ 
&\quad
+\left(\frac{\alpha_s^0}{2 \pi} \right)^2\mathrm{Split}^{(2)}_{P^h}(i^{h_i}, j^{h_j}) 
+ \mathcal{O}\left((\alpha_s^0)^3\right).
\end{split}\end{align}
We can then expand both sides of \eqref{eq:AllOrdersSplitting}, and we are particularly
interested in the order $(\alpha^0_s)^2$ contributions:
\begin{align}
    \bar{\mathcal{A}}^{(2)}(i^{h_i}, j^{h_j}, \ldots) \xrightarrow{i || j} 
    \sum_{h} \Big( &\mathrm{Split}^{(2)}_{P^h}(i^{h_i}, j^{h_j}) 
    \bar{\mathcal{A}}^{(0)}(\overline{P}^{-h}, \ldots) + 
    \mathrm{Split}^{(1)}_{P^h}(i^{h_i}, j^{h_j}) 
    \bar{\mathcal{A}}^{(1)}(\overline{P}^{-h}, \ldots) \nonumber\\
   	&+ \mathrm{Split}^{(0)}_{P^h}(i^{h_i}, j^{h_j}) 
   	\bar{\mathcal{A}}^{(2)}(\overline{P}^{-h}, \ldots) \Big).
  \label{eq:twoLoopSplitting}
\end{align}
It can be useful to understand this equation graphically, as presented in
figure \ref{fig:twoLoopSplitting}. As in \cref{eq:AllOrdersSplitting},
if the right-hand side vanishes, then the left-hand side is regular in the $i||j$ limit.

In the context of the three-photon production amplitudes at hand, many of the
collinear limits are equivalent due to the Bose symmetry of the final state. As the process is
unordered, the limits can be categorized according to the particles which become
collinear---either $(\gamma, \gamma)$, $(q, \overline{q})$, 
$(q,\gamma)$ or $(\overline{q},\gamma)$. 
These can then be further classified according to the helicity of the
involved particles. Our approach to check that our results satisfy 
\cref{eq:twoLoopSplitting} is to evaluate both sides of the equation
numerically. 
Note that this also tests the pentagon functions in near-singular
regions of phase-space.
To check their numerical behaviour, we constructed differential 
equations for the pentagon functions from their analytic representation,
which we then solved numerically to high precision using \cite{Hidding:2020ytt}. 
In the following, we discuss each of the different
collinear configurations. Before doing so, we note that we have
performed the same checks on the one-loop five-point amplitudes up to
order $\epsilon^2$. 
This allows one to check the implementation of the one-loop splitting amplitudes up to $\mathcal{O}(\epsilon^2)$ (see \cref{sec:1lSA} below), 
and is also a non-trivial check of the $\mathcal{O}(\epsilon)$ and
$\mathcal{O}(\epsilon^2)$ contributions to the
one-loop amplitudes which we have computed.

\paragraph{$(q, \overline{q})$ limit:} 
This limit is not accessible in the physical region for three-photon production
at hadron colliders, see \cref{eq:physregion}. For this reason equation
\cref{eq:twoLoopSplitting} cannot be checked in the $(q, \overline{q})$ limit.

\paragraph{$(\gamma, \gamma)$ limit:}
The only splitting amplitude that can contribute is
$\gamma^{h_1} \rightarrow \gamma^{h_2} \gamma^{h_3}$ which vanishes to all orders
for any helicity. It then follows that the amplitudes should be regular when any of 
the photons become collinear, independently of the helicity. We have verified that 
this is indeed the case for our amplitudes.

\paragraph{$(q, \gamma)$ limit:}
For the amplitude $\bar{\mathcal{A}}_{+++}$, given the fact that all the photons have 
the same helicity, there is a single limit to check corresponding to the splitting
$q\rightarrow \gamma^{+} q$. We note that the four-point amplitudes on the right-hand
side of \cref{eq:twoLoopSplitting} would then have a vanishing tree, and the
checks thus only involve the tree-level and one-loop splitting amplitudes.
For the amplitude $\bar{\mathcal{A}}_{-++}$, there
are two different limits to check, $q\rightarrow \gamma^{-} q$ 
and $q\rightarrow \gamma^{+} q$. In the first case, only
tree-level and one-loop splitting amplitudes are needed. In the latter case, 
two-loop splitting amplitudes also contribute. For the two cases
that only require up to one-loop splitting amplitudes, we checked 
\cref{eq:twoLoopSplitting} numerically, using the expressions
given in \cref{sec:1lSA} below. For the case that involves the two-loop
splitting amplitudes, we performed a different type of check. We first note
that loop-level splitting amplitudes, when normalized to the corresponding tree-level,
depend only on the vanishing $s_{ij}$ and the energy fraction $z$
 (see e.g.~\cref{eq:1lSA1,eq:1lSA2} below).
We can then consider two different collinear phase-space points which have 
the same values of $s_{ij}$
and $z$ but with all other kinematic parameters different. By evaluating our
amplitudes on these two points we can verify that the amplitudes are consistent
with the expected dependence of the normalized two-loop
splitting amplitude.

\paragraph{$(\bar{q}, \gamma)$ limit:} The situation is exactly the same as for 
the $(q, \gamma)$ limit.

\subsection{One-loop Photon Splitting Amplitudes}\label{sec:1lSA}

We write the splitting amplitudes as
\begin{equation}
\mathrm{Split}^{(l)}_{P_h}(i^{h_i}, j^{h_j})=
\rho^{(l)}(i^{h_i}, j^{h_j}) \mathrm{Split}^{(0)}(i^{h_i}, j^{h_j}).
\end{equation}
The $\rho^{(l)}(i^{h_i}, j^{h_j})$ are transcendental functions of
$s_{ij}$ and $z$, see \cref{eq:colLimZ}. Furthermore, they depend
on the scheme-defining parameter $D_s$, which is equal to $D$ in the HV scheme
but is kept generic here. Most of our checks only depend on the one-loop
splitting amplitudes, which can be obtained from the one-loop QCD splitting
amplitudes \cite{Bern:1999ry} with color-algebra manipulations. We first define
\begin{equation}
r_\Gamma(\epsilon) = 
\frac{1}{(4\pi)^\epsilon}\frac{\Gamma(1 - \epsilon)^2 \Gamma(1 + \epsilon)}
{\Gamma(1 - 2 \epsilon)}\,,
\end{equation}
and 
\begin{equation}
F(z,s_{ij},\epsilon)=r_\Gamma(\epsilon) \left(\frac{- s_{ij}}{\mu^2} \right)^{-\epsilon}
\frac{1}{\epsilon^2}
\sum_{m=1}^\infty \epsilon^m \mathrm{Li}_m\left(\frac{-z}{1-z}\right) \,.
\end{equation}
We then find that the ``distinct-helicity'' splitting amplitudes are
\begin{equation}\label{eq:1lSA2}
\rho_{q^+}^{(1)}(i_\gamma^+, j_q^-) = \rho_{q^-}^{(1)}(i_\gamma^-, j_q^+)
= F(z,s_{ij},\epsilon)\,,
\end{equation}
and the ``like-helicity'' splitting amplitudes are
\begin{equation}\label{eq:1lSA1}
\rho_{q^+}^{(1)}(i_\gamma^-, j_q^-) = \rho_{q^-}^{(1)}(i_\gamma^+, j_q^+) = 
F(z,s_{ij},\epsilon)-z\,r_\Gamma(\epsilon) 
\left(\frac{- s_{ij}}{\mu^2} \right)^{-\epsilon} 
\frac{D_s - 2}{4 (1 - \epsilon) (1 - 2 \epsilon)} \,.
\end{equation}

\bibliography{main.bib}

\end{document}